\documentclass[prb,twocolumn,showpacs]{revtex4}
\usepackage{color,soul}
\usepackage{amsmath}
\usepackage{dcolumn}
\usepackage{graphicx}
\usepackage{bm}
\usepackage{amssymb}
\usepackage{subfigure}
\begin{document}

\title{Strong relevance of Zinc impurity in the spin-$\frac{1}{2}$ Kagome quantum antiferromagnets: a variational study}
\author{Jianhua Yang and Tao Li}
\affiliation{Department of Physics, Renmin University of China, Beijing 100872, P.R.China}

\begin{abstract}
Copper hydroxyhalide materials herbertsmithite ZnCu$_{3}$(OH)$_{6}$Cl$_{2}$ and Zn-barlowite ZnCu$_{3}$(OH)$_{6}$FrBr are thought to be the best realizations of the spin-$\frac{1}{2}$ Kagome quantum antiferromagnetic Heisenberg model and are widely believed to host a spin liquid ground state. However, the exact nature of such a novel state of matter  is still under strong debate, partly due to the complication related to the occupation disorder between the Zinc and the Copper ions in these systems. In particular, recent nuclear magnetic resonance measurements indicate that the magnetic response of the Kagome plane is significantly spatial inhomogeneous, even though the content of the misplaced Zinc or Copper ions is believed to be very small. Here we use extensive variational optimization to show that the well known $U(1)$-Dirac spin liquid state is extremely sensitive to the introduction of the nonmagnetic Zinc impurity in the Kagome plane. More specifically, we find that the Zinc impurities can significantly reorganize the local spin correlation pattern around them and induce strong spatial oscillation in the magnetic response of the system. We argue that this is a general trend in highly frustrated quantum magnet systems, in which the nonmagnetic impurity may act as strongly relevant perturbation on the emergent resonating valence bond structure in their spin liquid ground state. We also argue that the strong spatial oscillation in the magnetic response should be attributed to the free moment released by the doped Zinc ions and may serve as the smoking gun evidence for the Dirac node in the $U(1)$ Dirac spin liquid state on the Kagome lattice. 
\end{abstract}

\maketitle
\section{Introduction}
The spin-$\frac{1}{2}$ Kagome antiferromagnet with nearest-neighboring Heisenberg exchange coupling(NN-KAFH) is widely believed to host quantum spin liquid ground state\cite{Elser,Chalker,Leung,Young,Lecheminant,Sindzingre,Nakano,Lauchli,Series,Vidal,Yan}. However,  even after the extensive study of the last three decades it is still elusive what is the exact nature of such a novel state of matter. While most variational studies suggest a gapless $U(1)$ Dirac spin liquid state\cite{Hastings,Ran,Iqbal1,Iqbal2,Iqbal3,Liao,He,Jiang,Zhu}, a gapped $Z_{2}$ spin liquid state has also been proposed by many studies\cite{Jiang1,Yan,Depenbrock,Jiang2,Kolley,Gong,Wen}. Adding to the weirdness is the observation of the exponential increase in the number of spin singlet excitations below the tiny triplet gap in exact diagonalization studies of the model on small cluster\cite{Singlet,Mila,Auerbach}. These problems have motivated several theoretical proposal claiming that the spin-$\frac{1}{2}$ NN-KAFH model may sit at or be very close to a quantum critical point\cite{Poilblanc,QCP1,QCP2,Tao1,Tao2}, where two or even a massive number of phases meet\cite{Sheng,He1,Changlani}. Perturbation away from such a singular point in the space of model parameter is thus expected to be highly nontrivial.

Indeed, the NN-KAFH model is special already at the classical level in the sense that it is fully frustrated\cite{Tao2}. More specifically, the lowest band of the Fourier transform of the exchange coupling is exactly flat on the Kagome lattice and the classical ground state of the system has no preferred classical ordering pattern at all. The massive degeneracy in the classical ground state provides a natural way to connect multiple competing phases. At the quantum level, it is found that the resonating valence bond(RVB) description of the spin liquid state on the Kagome lattice is also profoundly affected by the unique flat band physics on the Kagome lattice\cite{Tao2}. More specifically, the mapping between the RVB mean field ansatz and the physical spin liquid state becomes non-injective around the $U(1)$ Dirac spin liquid state, which is the best known variational ground state of the model. It is found that such a state can be generated from a continuous family of gauge inequivalent mean field ansatz with very different mean field spectrum. The conceptual link between the mean field excitation picture and the physical excitation spectrum is thus totally lost\cite{Tao3}.

The situation on the experimental side is by no means clearer. The Copper hydroxyhalide materials herbertsmithite ZnCu$_{3}$(OH)$_{6}$Cl$_{2}$ and Zn-barlowite ZnCu$_{3}$(OH)$_{6}$FrBr are claimed to be the best realizations of the spin-$\frac{1}{2}$ KAFH model\cite{Mendels,Helton,Shi}. In both materials, the spin-$\frac{1}{2}$ Cu$^{2+}$ ions form perfect Kagome planes with an antiferromagnetic exchange coupling as high as $J\approx 190 \mathrm{K}$ between nearest neighboring sites but remain paramagnetic down to a temperature as low as $10^{-4}J$. Debate on the nature of the low energy excitation spectrum of these Kagome materials starts right after their discoveries\cite{Han,Fu,Shi,Khunita}. Inelastic neutron scattering measurement on hebertsmithite ZnCu$_{3}$(OH)$_{6}$Cl$_{2}$ finds that the spin fluctuation spectrum of the system is characterized by a broad and featureless continuum above 2 meV.  Below 2 meV, a broad peak emerges in momentum space around the M point of the Brillouin zone\cite{Han}. If we attribute this low energy peak to the Cu$^{2+}$ impurity spins occupying the Zinc site out of the Kagome plane, a fully gapped spectrum would be inferred for the intrinsic spin fluctuation of the Kagome planes. Such a scenario is later supported by NMR measurement on both the herbertsmithite ZnCu$_{3}$(OH)$_{6}$Cl$_{2}$ and Zn-barlowite ZnCu$_{3}$(OH)$_{6}$FrBr\cite{Fu,Shi}, in which a spin gap of $\Delta\approx 0.03J \sim 0.07J$ is inferred. However, it is puzzling why the low energy spectral peak only appear around the M point if it is indeed caused by the fluctuation of local impurity spins\cite{Tao3}. A recent NMR measurement on hebertsmithite ZnCu$_{3}$(OH)$_{6}$Cl$_{2}$ shows on the other hand that the physics in the kagome planes is inhomogeneous and claims that the intrinsic spin fluctuation spectrum of the Kagome plane is gapless\cite{Khunita,Mendels1}. 

The controversy over the spin dynamics of these Kagome materials can be largely attributed to the complication related to the disorder effect in these systems\cite{Han,Freedman,Smaha,Wang1,Yuan1}. The occupation disorder between the Cu$^{2+}$  and the Zn$^{2+}$ ions, which have the same charge and very similar radius, is the most discussed origin of disorder in these Kagome materials. A site-selective XRD measurement indicates that there is about $15\%$($5\%$) excess Cu$^{2+}$ impurity spin occupying the nonmagnetic Zn$^{2+}$ site between the Kagome layers in the herbertsmithite ZnCu$_{3}$(OH)$_{6}$Cl$_{2}$(Zn-barlowite ZnCu$_{3}$(OH)$_{6}$FrBr)\cite{Freedman,Smaha}. On the other hand, the upper bound of nonmagnetic Zn$^{2+}$ ion occupying the Cu$^{2+}$ site in the Kagome plane is estimated to be less than $1\%$ in both systems. Recent NMR measurements indicate that the small amount of out-of-plane Cu$^{2+}$ impurity spin can significantly alter the spin dynamics of the Kagome plane. For example, through an inverse Laplace transform analysis of the spin relaxation rate on the Zn-barlowite ZnCu$_{3}$(OH)$_{6}$FrBr, the authors of Ref.[\onlinecite{Wang1}] conclude that at temperature far below the exchange coupling only about $60\%$ of the spin in the Kagome plane are involved in the formation of spin singlet pair, with significant spatial inhomogeneity in the size of their excitation gap. Using a more recent two dimensional NMR data analysis scheme on the Knight shift and relaxation rate data of the Zn-barlowite ZnCu$_{3}$(OH)$_{6}$FrBr, the author of Ref.[\onlinecite{Yuan1}] conclude that the emergent spin polarized domain induced by the interlayer Cu$^{2+}$ impurity spin account for $60\%$ of volume at low temperature, even though the content of such impurity spin is estimated to be only about $5\%$.   

The strong relevance of disorder effect in a highly frustrated magnet such as the Kagome antiferromagnet is in a sense naturally expected. At the semiclassical level, the introduction of a local impurity will relieve the local frustration between the exchange couplings, which may again trigger the rebalance between frustrated couplings in more remote regions. This effect is expected to be especially relevant when the model is fully frustrated at the semiclassical level.  At the quantum level, since the RVB structure emerges on the background of delicately balanced exchange couplings, any perturbation that break such subtle balance is expected to have dramatic effect on such an emergent structure. The introduction of local impurity in such highly frustrated background is thus expected to have highly nontrivial consequences. Such an expectation is indeed supported by several theoretical studies on the impurity effect on the Kagome antiferromagnet, of either classical or quantum version.\cite{Henley,Mila1,Mila2,Motrunich,Moessner1,Moessner2,Patil}

With these considerations in mind, we have performed a systematic variational Monte Carlo(VMC) study on the ground state and magnetic response of a disordered spin-$\frac{1}{2}$ KAFH model within the Fermionic RVB picture. To be more specific, we study the spin-$\frac{1}{2}$ KAFH model on a finite cluster with a pair of nonmagnetic Zn$^{2+}$ impurity occupying the Cu$^{2+}$ site within the Kagome plane. The reason to choose such a specific disorder model is twofold. First, while the content of Zn$^{2+}$ impurity in the Kagome plane is much lower than that of the Cu$^{2+}$ impurity out of the Kagome plane in real materials, the former is expected to have much stronger effect on the properties of the Kagome plane. Second, the nonmagnetic impurity within the Kagome plane can be simply modeled as a vacancy in the plane. This is very different from the case of out-of-plane Cu$^{2+}$ impurity, the modeling of which need the introduction of many more phenomenological parameters that we have no control.

Through large scale variational optimization of the spin-$\frac{1}{2}$ KAFH model, we find that the Zinc ions will significantly reorganize the local spin correlation pattern around them. The most significant effect of the doped Zinc ion manifests itself in the strong spatial inhomogeneity in the magnetic susceptibility of the system at low field. We find that the magnetization of the Zinc-doped system exhibits strong spatial oscillation with a dominant wave vector of $\frac{1}{2}\mathbf{b}_{1}$ or its symmetry equivalents, which are nothing but the momentum difference between the two Dirac nodes of the $U(1)$ Dirac spin liquid state on the Kagome lattice. The amplitude of such oscillation is found to be even larger than the average magnetization expected for the pure system. We argue that such strong spatial inhomogeneity in the magnetic response should be attributed to the free moment released by the doped Zinc ions and may serve as smoking gun evidence for the Dirac node in the $U(1)$ spin liquid state on the Kagome lattice. Our variational results are consistent with the observations of recent NMR measurements on the Kagome materials\cite{Wang1,Yuan1} and suggest the strong relevance of disorder effect in such highly frustrated quantum magnet systems.

The paper is organized as follows. In the next section, we introduce the Zinc-doped spin-$\frac{1}{2}$ KAFH model and its variational description in terms of the Fermionic RVB theory. In Sec.III, we introduce the physical quantities that will be studied on such a disordered system. In particular, we will discuss how to calculate the local susceptibility and the center of gravity of the local spin fluctuation spectrum within the Fermionic RVB theory. In Sec.IV, we will present the numerical results of the variational calculation. The last section of the paper is devoted to the conclusion of our results and a discussion on some general issues about disorder effect in highly frustrated quantum magnets.

\section{The Zinc-doped spin-$\frac{1}{2}$ KAFH model and its variational description}
The Zinc-doped spin-$\frac{1}{2}$ KAFH model studied in this work has the Hamiltonian
\begin{equation}
H_{J}=J\sum_{\langle i,j\rangle\neq i_{0},i_{1}}\mathbf{S}_{i}\cdot\mathbf{S}_{j}.
\end{equation}
The sum is over nearest-neighboring bonds of a finite Kagome cluster with the two sites $i_{0}$ and $i_{1}$ occupied by the doped Zinc ions excluded. In the majority of this work, we will concentrate on a $12\times12\times3$ Kagome cluster with periodic boundary condition along both the $\mathbf{a}_{1}$ and the $\mathbf{a}_{2}$ direction, an illustration of which is shown in Fig.1. In the following, we will set $J=1$ as the unit of energy.

\begin{figure}
\includegraphics[width=8.5cm]{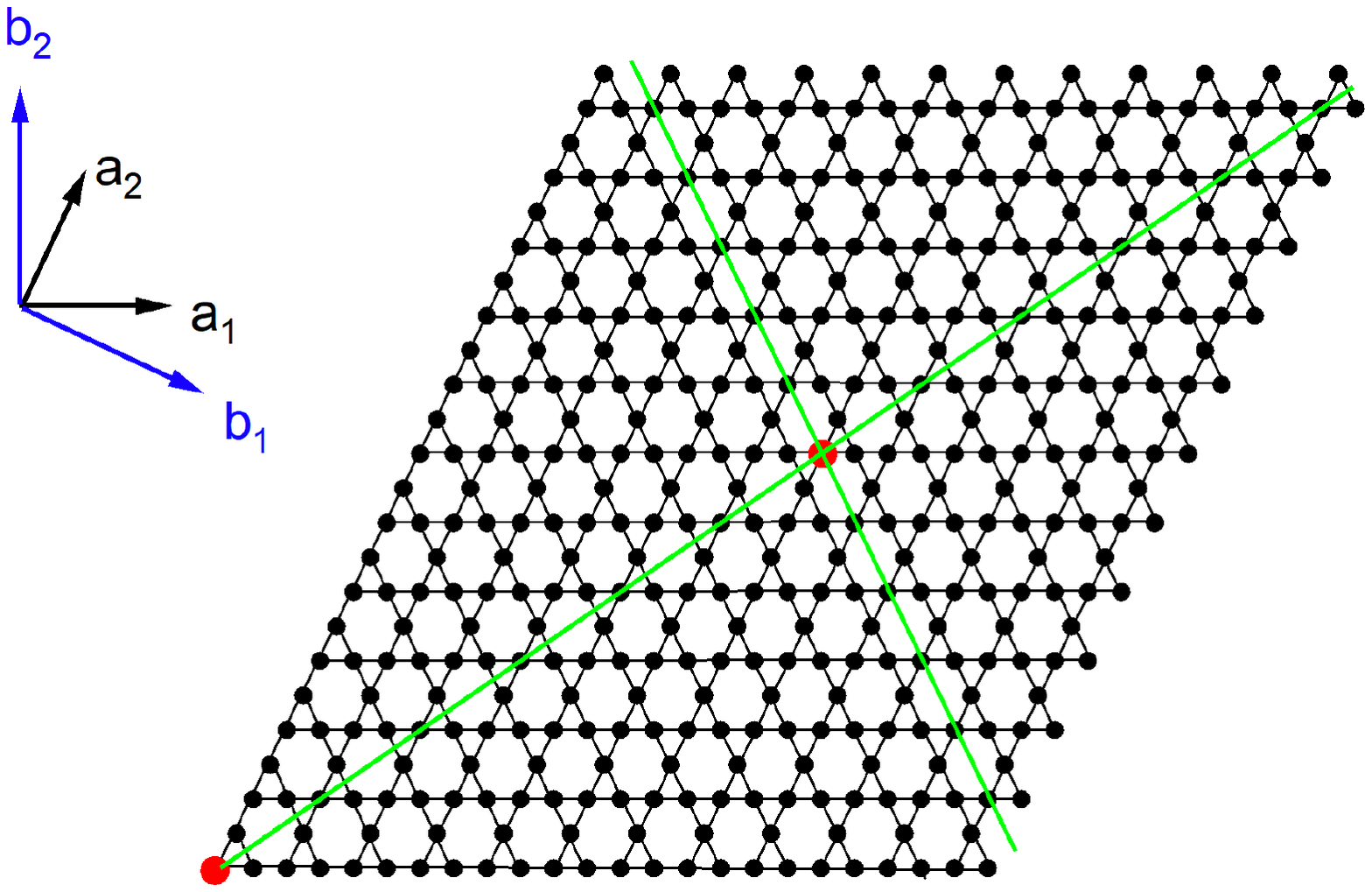}
\caption{Illustration of the finite Kagome cluster on which our variational calculation is done. Periodic boundary condition is assumed along the direction of $\mathbf{a}_{1}$ and $\mathbf{a}_{2}$ so that the cluster forms a two dimensional torus. Here $\mathbf{a}_{1}$ and $\mathbf{a}_{2}$ are the two basis vectors of the Kagome lattice and  $\mathbf{b}_{1}$ and $\mathbf{b}_{2}$ are the corresponding reciprocal vectors. The red dots at the corner and the center of the cluster denote the position of doped Zinc ions, which act as vacancy in the Kagome lattice. Most of our computations are done on a $N_{c}=12\times12\times3$ cluster, which has $N=N_{c}-2=430$ lattice site. The cluster is invariant under the reflection about the two thick green lines through the two doped Zinc ions.}
\end{figure}

To describe the ground state of the system in the RVB scheme, we represent the spin operator as 
\begin{equation}
\mathbf{S}=\frac{1}{2}\sum_{\alpha,\beta}f^{\dagger}_{\alpha}\bm{\sigma}_{\alpha,\beta}f_{\beta}
\end{equation}
in which $f_{\alpha}$ is a Fermionic slave particle operator, $\bm{\sigma}$ is the Pauli matrix. The representation becomes exact when the following constraint
\begin{equation} 
\sum_{\alpha}f^{\dagger}_{\alpha}f_{\alpha}=1
\end{equation}
is satisfied. The Fermionic RVB state that we will adopt to describe the ground state of the model is generated from Gutzwiller projection of  the mean field ground state of the following BCS Hamiltonian
\begin{equation}
H_{MF}=\sum_{i,j}\psi_{i}^{\dagger}U_{i,j}\psi_{j}.
\end{equation}
Namely
\begin{equation}
|f-RVB\rangle=P_{G}|MFG\rangle
\end{equation}
in which $P_{G}$ is the Gutzwiller projection operator removing the doubly occupied configuration from the mean field ground state $|MFG\rangle$.
Here 
\begin{equation}
\psi_{i}=\left(\begin{array}{c}f_{i,\uparrow}\\f^{\dagger}_{i,\downarrow}\end{array}\right),
\end{equation}
is a two component spinor made of the spinon operator,
\begin{equation} 
U_{i,j}=\left(\begin{array}{cc}\chi_{i,j} & \Delta^{*}_{i,j} \\\Delta_{i,j} & -\chi^{*}_{i,j}\\ \end{array}\right)
\end{equation}
is the matrix form of the RVB order parameter.  $\chi_{i,j}$ and $\Delta_{i,j}$ denote the RVB parameters in the hopping and pairing channel and are used as variational parameters in our study. The mean field Hamiltonian $H_{MF}$ is usually called a mean field ansatz of the RVB state. In our study, we will concentrate on mean field ansatz with real $\chi_{i,j}$ and real $\Delta_{i,j}$ between nearest neighboring sites. 

The RVB state so constructed is invariant under a $SU(2)$ gauge transformation of the form
\begin{eqnarray} 
\psi_{i}&\rightarrow &G_{i} \psi_{i}\nonumber\\
U_{i,j}&\rightarrow& G_{i}U_{i,j}G^{\dagger}_{j}\nonumber\\
\end{eqnarray}
Here $G_{i}$ is a site-dependent $SU(2)$ matrix\cite{Wen1}. In the literature, a mean field ansatz with only nonzero $\chi_{i,j}$ is called a $U(1)$ ansatz. A mean field ansatz that can not be taken into the $U(1)$ form under any $SU(2)$ gauge transformation is called a $Z_{2}$ ansatz.

In the absence of the Zinc impurity, it is generally believed that the best variational ground state of the spin-$\frac{1}{2}$ KAFH model is generated by the following $U(1)$ Dirac mean field ansatz
\begin{equation}
\chi_{i,j}=s_{i,j}
\end{equation}
with a variational energy of $E\approx-0.4286$ per site\cite{Ran}. Here $\chi_{i,j}\neq0$ only between nearest neighboring sites, $s_{i,j}=\pm1$ is a sign factor introduced to guarantee that a gauge flux of $\pi$ is enclosed in each elementary hexagon of the Kagome lattice and that zero flux is enclosed in each elementary triangle of the Kagome lattice. A realization of ansatz is illustrated in Fig.2. 

\begin{figure}
\includegraphics[width=10cm]{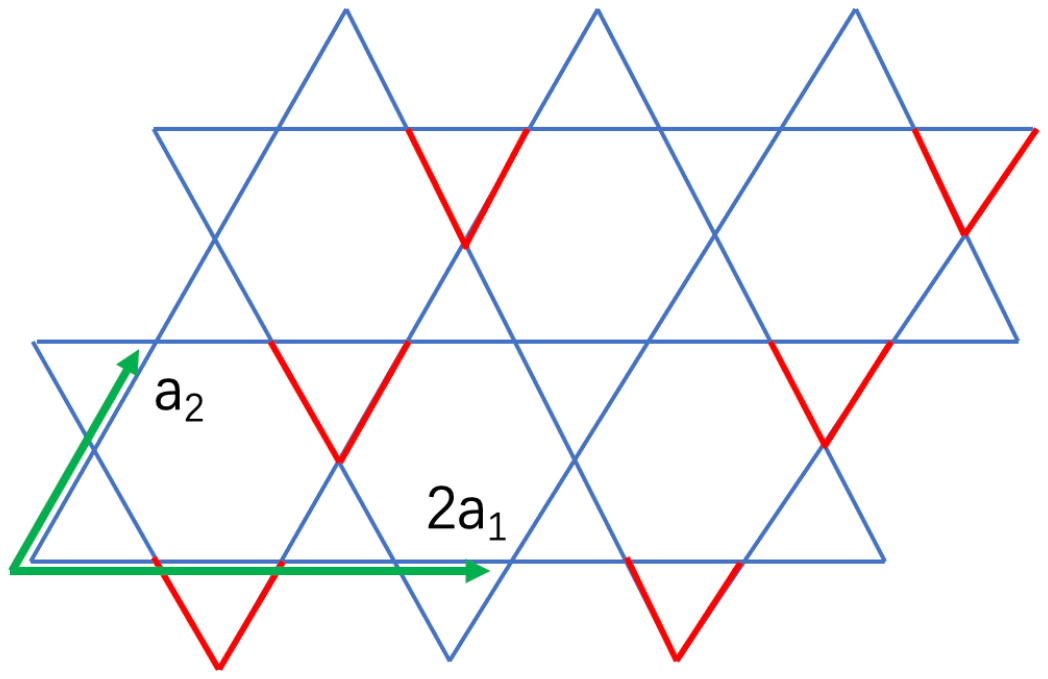}
\caption{A realization of the mean field ansatz of the $U(1)$ Dirac spin liquid state on the Kagome lattice. The unit cell of the spinon is doubled in the direction of $\mathbf{a}_{1}$ as a result of the $\pi$ flux enclosed in each elementary hexagon of the Kagome lattice. $s_{i,j}=-1$ on blue bonds and $s_{i,j}=1$ on red bonds.}
\end{figure}

In our study, we will adopt the $U(1)$ Dirac ansatz as our initial guess for the variational ground state of the Zinc-doped system(with the RVB order parameter on those bonds connected to the Zinc site set to zero). We note that the mapping between the mean field ansatz and the RVB state is non-injective around the $U(1)$ Dirac ansatz as a result of the unique flat band physics on the Kagome lattice\cite{Tao2}. Continuous family of gauge inequivalent RVB mean field ansatz, of either the $U(1)$ or the $Z_{2}$ form may correspond to very close RVB state\cite{Tao3,Lu}. For this reason, we will consider variational ansatz of both the $U(1)$ and the $Z_{2}$ form.

More specifically, the $U(1)$ variational ansatz we consider takes the form of
\begin{equation}
H_{MF}=\sum_{\langle i,j\rangle,\sigma}\chi_{i,j}f_{i,\sigma}^{\dagger}f_{j,\sigma}+\sum_{i,\sigma}\mu_{i}f_{i,\sigma}^{\dagger}f_{i,\sigma}.
\end{equation}
in which $\chi_{i,j}=\chi_{j,i}=real$ is the hopping parameter between site $i$ and $j$, $\mu_{i}=real$ is a local chemical potential. These are the variational parameters of the problem. On the $12\times12\times3$ Kagome cluster, there are in total 1286 real variational parameters to be optimized. Using reflection symmetry of the finite Kagome cluster about the two green lines in Fig.1, we can reduce the number of variational parameter in the $U(1)$ ansatz to 333. In the Ising basis, the RVB state so generated takes the form of
\begin{equation}
|f-RVB\rangle=\sum_{\{\sigma_{i}\}}\Psi(\{\sigma_{i}\})|\sigma_{1},....,\sigma_{N}\rangle
\end{equation} 
in which 
\begin{equation}
|\sigma_{1},....,\sigma_{N}\rangle=\prod_{k=1,N/2}f^{\dagger}_{i_{k},\uparrow}f^{\dagger}_{j_{k},\downarrow}|0\rangle
\end{equation}
is the Ising basis generated by the spinon operator, $|0\rangle$ is the Fermion vacuum. $i_{k}$ and $j_{k}$ are the coordinate of the $k$-th up spin and the $k$-th down spin. The wave function amplitude $\Psi(\{\sigma_{i}\})$ is given by
\begin{equation}
\Psi(\{\sigma_{i}\})=\mathrm{Det}[\bm{\Phi}_{\uparrow}]\times\mathrm{Det}[\bm{\Phi}_{\downarrow}]
\end{equation}
in which $\mathrm{Det}[\bm{\Phi}_{\uparrow}]$ and $\mathrm{Det}[\bm{\Phi}_{\uparrow}]$ are the Slater determinant of the up spin Fermions and the down spin Fermions corresponding to the filling of the lowest $N/2$ eigenstates of $H_{MF}$.

The $Z_{2}$ variational ansatz we consider takes the form of
\begin{eqnarray}
H_{MF}&=&\sum_{\langle i,j\rangle,\sigma}\chi_{i,j}f_{i,\sigma}^{\dagger}f_{j,\sigma}+\sum_{i,\sigma}\mu_{i}f_{i,\sigma}^{\dagger}f_{i,\sigma}\nonumber\\
&+&\sum_{\langle i,j\rangle}\Delta_{i,j}(f_{i,\uparrow}^{\dagger}f_{j,\downarrow}^{\dagger}+h.c.)
\end{eqnarray} 
On the $12\times12\times3$ Kagome cluster, there are in total 2142 real variational parameters to be optimized. Using reflection symmetry of the finite Kagome cluster, we can reduce the number of variational parameter in the $Z_{2}$ ansatz to 553. In our calculation, we have adopted the following particle-hole transformation on the down-spin Fermions
\begin{equation}
f^{\dagger}_{i,\downarrow}\rightarrow\tilde{f}_{i,\downarrow}
\end{equation}
In the Ising basis, the RVB state so generated takes the form of
\begin{equation}
|f-RVB\rangle=\sum_{\{i_{1},...i_{N/2}\}}\tilde{\Psi}(i_{1},...,i_{N/2})\prod_{k=1,..,N/2}f^{\dagger}_{i_{k},\uparrow}\tilde{f}^{\dagger}_{i_{k},\downarrow}|\tilde{0}\rangle
\end{equation}
in which 
\begin{equation}
|\tilde{0}\rangle=\prod_{i=1,..,N}f^{\dagger}_{i_{k},\downarrow}|0\rangle
\end{equation}
is the reference state with all site occupied by down spin Fermions. The wave function amplitude $\tilde{\Psi}(i_{1},...,i_{N/2})$ is given by 
\begin{equation}
\tilde{\Psi}(i_{1},...,i_{N/2})=\mathrm{Det}[\bm{\Phi}]
\end{equation}
Here $\mathrm{Det}[\bm{\Phi}]$ is the Slater determinant corresponding to the filling of the lowest $N$ eigenstates of the particle-hole transformed mean field Hamiltonian. 

We will mainly adopt the stochastic reconfiguration(SR) algorithm in the variational optimization\cite{Sorella}. In the SR algorithm, the variational parameters(denoted here as $\bm{\alpha}$) are updated as follows
 \begin{equation}
\bm{\alpha}\rightarrow \bm{\alpha}-\delta\ \mathbf{S}^{-1}\nabla E
\end{equation} 
in which $\nabla E$ is the gradient of the variational energy with respect to the variational parameters, $\mathbf{S}$ is a positive-definite and Hermitian matrix generated from the metric of the variational state and is given by
 \begin{equation}
 \mathbf{S}=\langle \nabla ln\Psi \nabla ln\Psi \rangle-\langle \nabla ln\Psi \rangle\langle \nabla ln\Psi \rangle
 \end{equation}
 $\delta$ is the step length which is usually chosen by trial and error. The SR method can be accelerated by the following self-learning trick\cite{Tao4}. Denoting the current search direction suggested by the SR method as $\mathbf{g}$(assumed to be normalized), then the step length $\delta$ is updated as follows
 \begin{equation}
 \delta\rightarrow \delta\times(1+\eta \ \mathbf{g}\cdot \mathbf{g}')
 \end{equation}  
 in which $\eta\in (0,1)$ is an acceleration factor, $\mathbf{g}'$ is the search direction in the previous step. To see the logic of the trick, let us set $\eta=1$. Then the step length will be doubled(suppressed to zero) if $\mathbf{g}$ is parallel(antiparallel) to $\mathbf{g}'$ and will be unchanged if  $\mathbf{g}$ is orthogonal to $\mathbf{g}'$. In practice, we should set the upper limit for the step length to avoid possible run away from true minimum. We should also set the lower limit for the step length to avoid possible false convergence. Very close to the minimum, the SR method may suffer from slow convergence or run away from the minimum. In this case, we adopt the BFGS algorithm to refine the solution we get from the SR method. The details on the implementation of the BFGS method can be found in Ref.[\onlinecite{Tao4}].

\section{The physical quantities and their variational calculation}
Beside the variational ground state energy, we are also interested in various other physical quantities in the variational study. To diagnose the influence of the Zinc impurity on the system, we have calculated the local spin correlation, the local spin susceptibility and the center of gravity of the local spin fluctuation spectrum around the doped Zinc ions.  

\subsection{The ground state properties}
\subsubsection{The local spin correlation}
The local spin correlation is spatially uniform and isotropic in the $U(1)$ Dirac spin liquid state, with $\langle \mathbf{S}_{i}\cdot\mathbf{S}_{j} \rangle\approx  -0.2143$ between nearest neighboring sites. Naively, one expects that the introduction of the pair of Zinc ions would just cut off the spin correlation on the 8 nearest neighboring bonds connected to the two sites occupied by doped Zinc ions. Correspondingly, one would expect that the ground state energy of the Zinc doped system to be about $8\langle \mathbf{S}_{i}\cdot\mathbf{S}_{j} \rangle\approx 4\times -0.4286=-1.7144$ higher than that of the pure system. 

However, the introduction of the doped Zinc ions will break the delicate balance between the highly frustrated exchange couplings on the Kagome lattice and will thus reorganize the local spin correlation pattern around them. In particular, on the triangle containing the doped Zinc ions, the remaining two spins will have large chance to form spin singlet. This will again trigger the reorganization of the spin correlation pattern in more remote regions, just as the situation in an avalanche event. If the system is nearly critical toward the formation of certain valence bond solid order, the introduction of the Zinc impurity would even induce divergent response in the local spin correlation pattern. Our calculation shows that the actual increase in the ground state energy by the doped Zinc ions is much lower than the naive broken-bond estimate we made above, indicating significant reorganization of the local spin correlation pattern around the Zinc ions. It is important to see how such reorganization would decay with distance.

 \subsubsection{The center of gravity of the local spin fluctuation spectrum}
To study the spin relaxation rate we need the spin fluctuation spectrum of the system, which are related to each other by
\begin{equation}
\frac{1}{TT_{1}} \propto  |A_{hf}|^{2} \frac{\chi^{''}(\omega_{0})}{\omega_{0}}
\end{equation}  
Here $A_{hf}$ is the hyperfine coupling constant between the nuclear spin and the electron spin, $\omega_{0}$ is the resonance frequency in the NMR measurement. In principle, we can calculate the full spin fluctuation spectrum within the dynamical variational theory scheme\cite{Tao5,Tao3}. However, without translational symmetry, the number of projected basis functions needed to perform such a dynamical variational calculation increases rapidly with the system size. For example, on our $N$=430 Kagome cluster, the number of projected basis functions is $N(N-1)/2=92450$, which is beyond our reach currently. In addition, the NMR relaxation rate is determined by the low energy limit of the spin fluctuation spectrum, which is inaccessible on a small cluster with only 430 sites. 

For these reasons, we turn to the calculation of a much easier dynamical quantity, the center of gravity of the local spin fluctuation spectrum. It is defined as
\begin{equation}
\langle \omega \rangle_{i,i}=\frac{\int d\omega \omega \chi^{''}_{i,i}(\omega)}{\int d\omega \chi^{''}_{i,i}(\omega)}
\end{equation} 
in which $\chi^{''}_{i,i}(\omega)$ denotes the fluctuation spectrum of the spin on site $i$. Using local spin sum rule, it can be shown that 
\begin{equation}
\langle \omega \rangle_{i,i}=\frac{1}{2}\frac{\langle[\mathbf{S}_{i},[H,  \mathbf{S}_{i}]] \rangle}{\langle \mathbf{S}_{i} \cdot \mathbf{S}_{i}\rangle}
\end{equation} 
It is then straightforward to show that for the spin-$\frac{1}{2}$ KAFH model
\begin{equation}
\langle \omega \rangle_{i,i}=-\frac{8}{3}\sum_{\delta}\langle \mathbf{S}_{i}\cdot\mathbf{S}_{i+\delta}\rangle
\end{equation}
in which $i+\delta$ denotes the nearest neighbor of site $i$ on the Kagome lattice. In the variational calculation, we replace the exact ground state in the above formula with the variational ground state.

While $\langle \omega \rangle_{i,i}$ only contains partial information on the local spin fluctuation spectrum, it may still be valuable for the analysis of the spatial inhomogeneity of the spin relaxation rate. More specifically, if we assume that the spectral form of $\chi^{''}_{i,i}(\omega)$ is approximately site independent(an approximation which becomes valid when the influence of the impurity is sufficiently local and we are sufficiently far away from it), $\langle \omega \rangle_{i,i}$ would be inversely proportional to the initial slope of $\chi^{''}_{i,i}(\omega)$, since the total spectral weight contained in $\chi^{''}_{i,i}(\omega)$ is quantized. $\langle \omega \rangle_{i,i}$ would thus be inversely proportional to the spin relaxation rate. As we will see in the next section, such an approximation fails at low energy even if we are far away from the Zinc site since the magnetic response from the Zinc induced free moment is spatially nonlocal.

\subsection{The local spin susceptibility}
Recent NMR measurement shows that significant spatial inhomogeneity can be generated in the magnetic response of the Kagome plane by small amount of magnetic impurity, in both the Knight shift and the spin relaxation rate channel. For example, while the content of the Cu$^{2+}$ ion occupying the out-of-plane Zn$^{2+}$ site is estimated to be only $5\%$ in the Zn-barlowite ZnCu$_{3}$(OH)$_{6}$FrBr, it is found that about $60\%$ volume of the system are involved in the emergence of spin polarized domains\cite{Yuan1}.

Here we calculate the local susceptibility as an equilibrium property within the Fermionic RVB scheme. We concentrate on the low field limit in which only a pair of spinons are excited. There are two ways to proceed the calculation. In the first approach, we first optimize the variational parameters in the ground state wave function. We then construct the variational wave function of the magnetized state by exciting a pair of spinons on the RVB mean field ground state before performing the Gutzwiller projection. The spatial distribution of the field induced magnetization is calculated on such a variational state. More specifically, we propose the following variational description for the magnetized state
\begin{equation}
|Mag\rangle=P_{G}\gamma^{\dagger}_{1,\uparrow}\gamma^{\dagger}_{2,\uparrow}|MFG\rangle
\end{equation}     
in which $\gamma^{\dagger}_{1,\uparrow}$ and $\gamma^{\dagger}_{2,\uparrow}$ denote the creation operator for the two lowest spinon excitations on the RVB mean field ground state. In case of the $U(1)$ RVB state, the pair of spinons are excited by the lowest particle-hole transition across the Fermi level. 

In the second approach, we directly optimize the variational parameters in $|Mag\rangle$ and then calculate the resultant magnetization. By so doing the relaxation of the RVB order parameter in the magnetization process is taken into account. We find that the spatial distribution of the magnetization is insensitive to such a relaxation process. By the way, we note that even for a pure system the magnetization is spatially uniform only when the closed shell condition is satisfied in $|Mag\rangle$. We find that the optimized variational state $|Mag\rangle$ in the presence of the doped Zinc ions always satisfy the closed shell condition. The spatial oscillation in the magnetization we find below from the optimized $|Mag\rangle$ should thus be attributed to the Zinc doping effect, rather than a finite size effect.

\section{Numerical results}
Now we present the numerical results on both the ground state properties and the magnetic susceptibility of the Zinc doped spin-$\frac{1}{2}$ KAFH model.

\begin{figure}
\includegraphics[width=8cm]{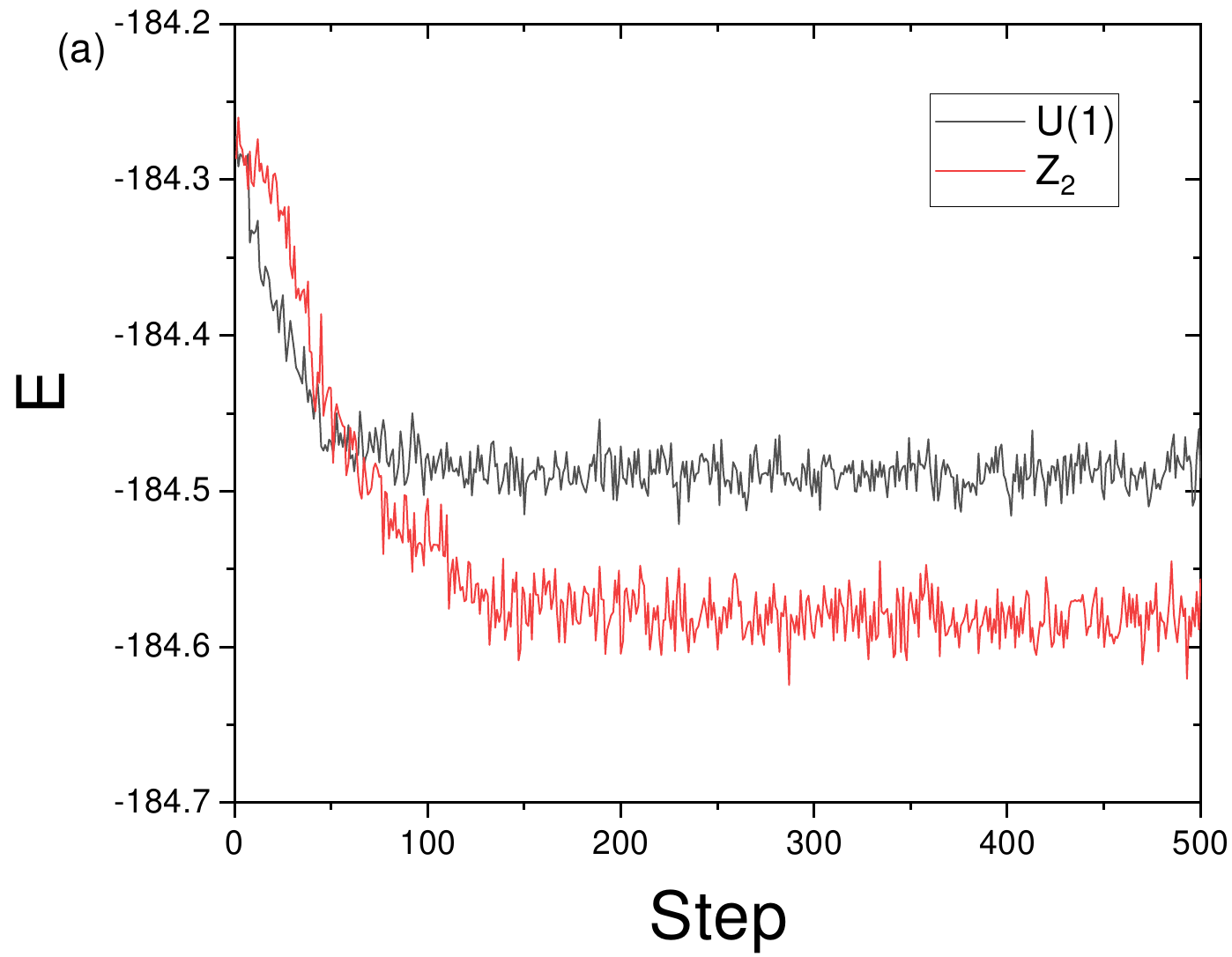}
\includegraphics[width=8cm]{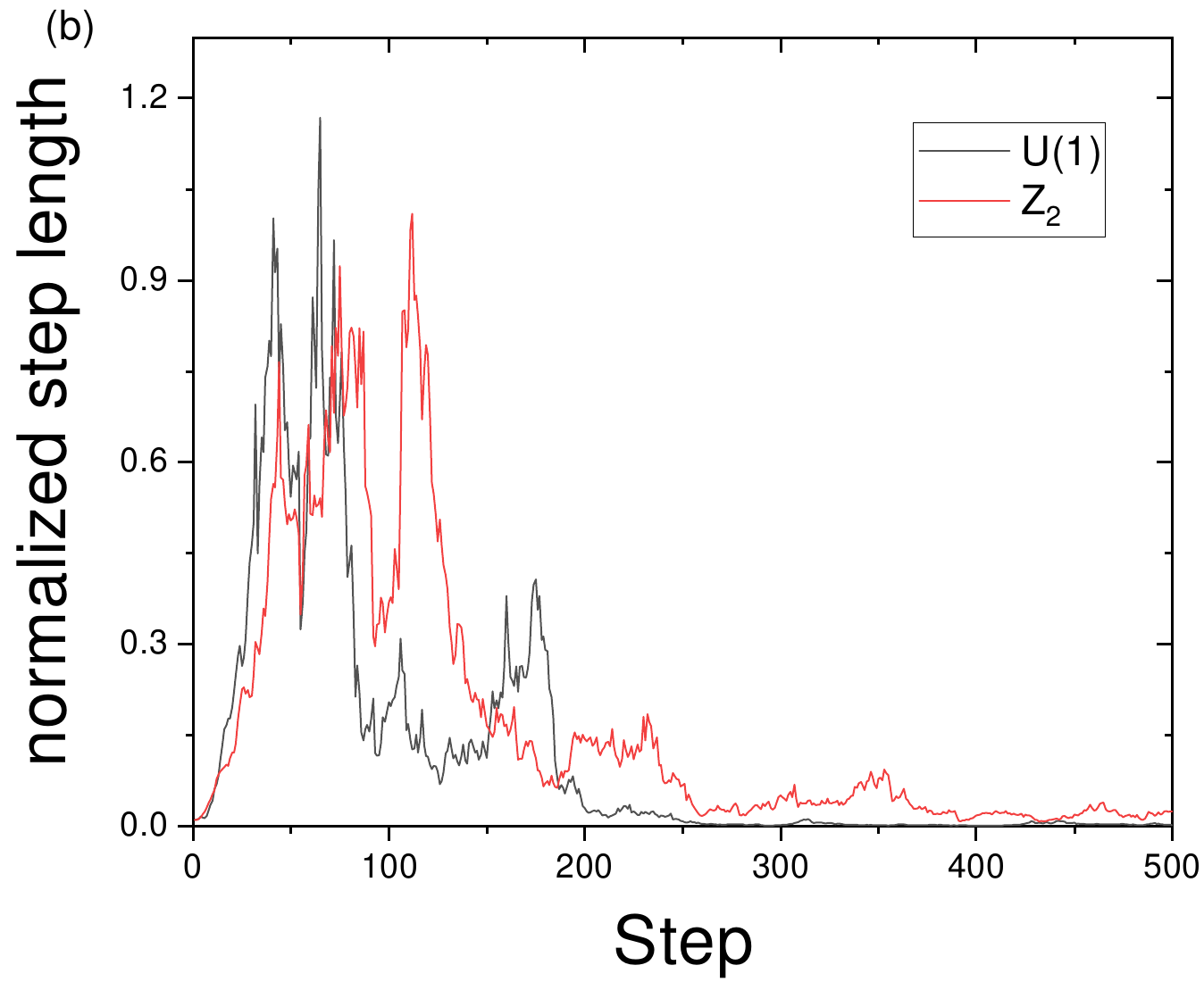}
\caption{(a)The convergence of the variational energy for the $U(1)$ and $Z_{2}$ state of the spin-$\frac{1}{2}$ KAFH model with a pair of doped Zinc ions on a $12\times12\times3$ cluster illustrated in Fig.1. The optimization is done with the SR method accelerated by the self-learning trick. The evolution of the step length during the optimization is shown in (b).}
\end{figure}

\subsection{The ground state properties}
We have performed variational optimization for both the $U(1)$ and the $Z_{2}$ state on the $12\times12\times3$ cluster shown in Fig.1 with the SR algorithm accelerated by the self-learning trick. As is shown in Fig.3, the variational ground state energy converges rapidly to $E_{U(1)}\approx -184.49$ and $E_{Z_{2}}\approx -184.59$ for the $U(1)$ and the $Z_{2}$ state respectively. Here we have adopted the reflection symmetry about the two green lines in Fig.1 to refine the optimization. The increase of the ground state energy with respect to that of the pure system(which is $E_{pure}\approx -185.15$) is thus about $E_{U(1)}-E_{pure}\approx-0.66$ and $E_{Z_{2}}-E_{pure}\approx-0.56$ for the $U(1)$ and the $Z_{2}$ state. Both of which are thus significantly (about a factor of three) smaller than the naive estimate of $-1.7144$ from the broken bond picture, implying significant reorganization of the local spin correlation pattern around the Zinc ions.
 
 \begin{figure}
\includegraphics[width=9cm]{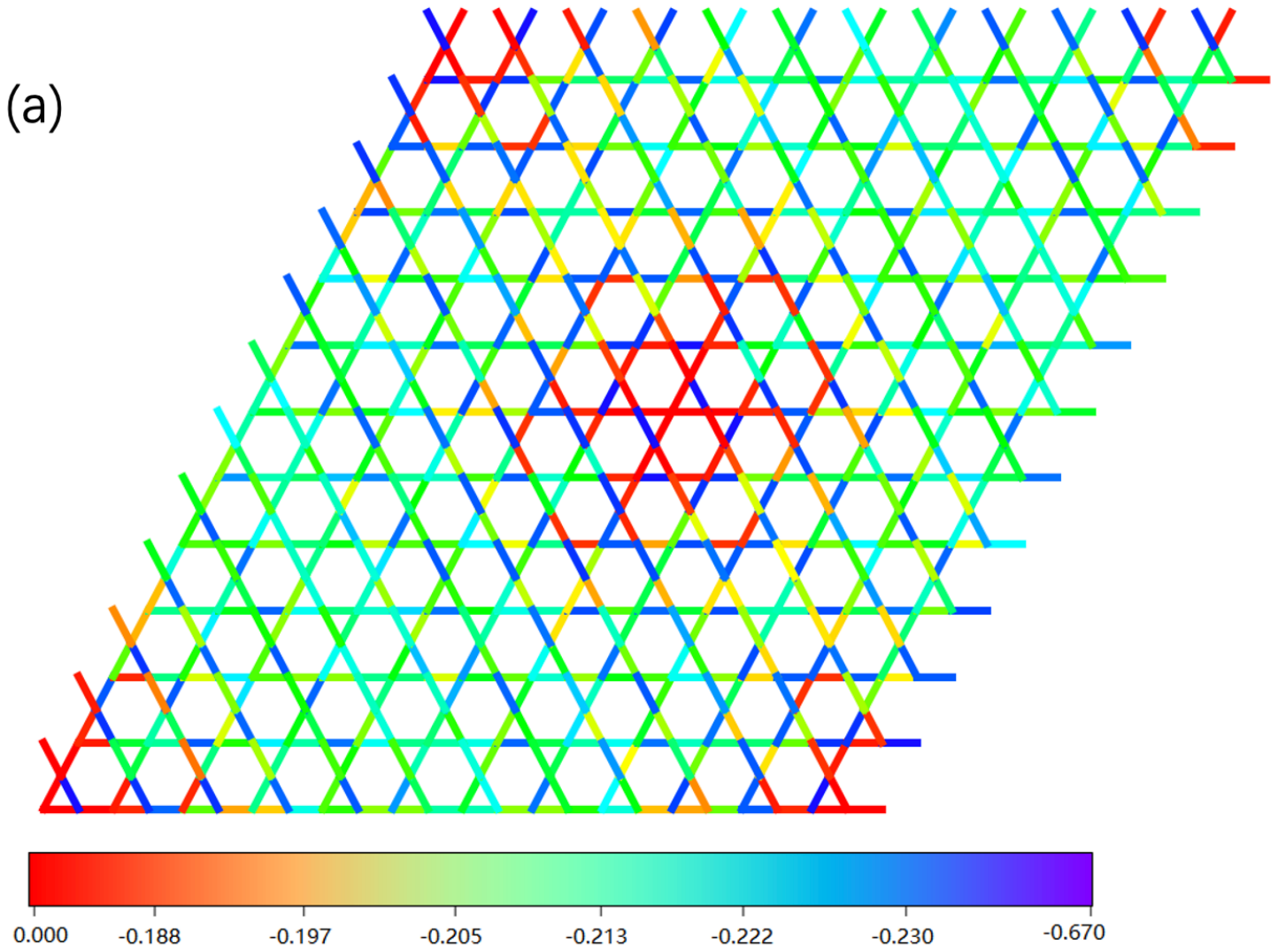}
\includegraphics[width=8cm]{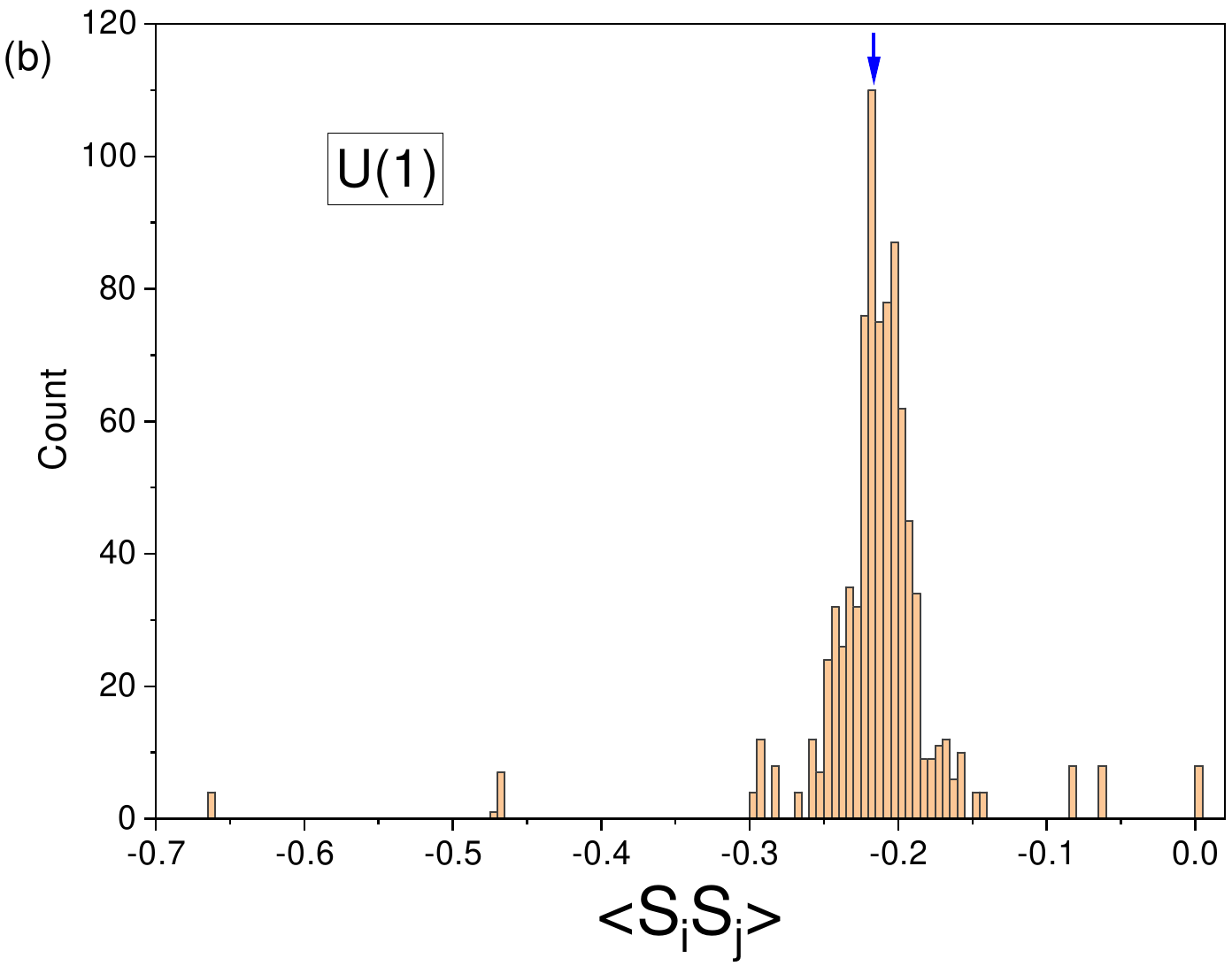}
\caption{(a)The local spin correlation pattern in the optimized $U(1)$ state. Note that the color scale is not uniform. (b)The distribution of the spin correlation between nearest neighboring sites. The blue arrow marks the value of nearest neighboring spin correlation in the pure spin-$\frac{1}{2}$ KAFH model.}
\end{figure}

The local spin correlation in the optimized $U(1)$ and the $Z_{2}$ state are shown in Fig.4 and Fig.5 and are found to be very similar with each other. The most prominent feature in the local spin correlation pattern is the strong singlet correlation between the two remaining spins on the triangle containing the Zinc ions. In more remote region, the local spin correlation exhibits oscillating behavior. More specifically, when a spin is involved in a strong bond with a nearest neighboring spin, its correlation with other nearest neighboring spins is suppressed. This is consistent with our qualitative discussion in the last section. 

Fig.4b and Fig.5b show the distribution in the value of the nearest neighboring spin correlation on the Kagome lattice. The spatial inhomogeneity of the local spin correlation in both states are quite significant. More specifically, in the optimized $U(1)$ state, the spin correlation on more than $35\%$ of nearest neighboring bonds differ from their value in the pure system by more than $10\%$. For the optimized $Z_{2}$ state, the spin correlation on more than $22\%$ of nearest neighboring bonds differ from their value in the pure system by more than $10\%$. The spatial inhomogeneity of the local spin correlation in the $Z_{2}$ state is thus slightly weaker than that in the $U(1)$ state.

We note that qualitatively similar behavior has been observed in the local spin correlation pattern obtained by exact diagonalization study on small Kagome clusters\cite{Mila1}. In particular, the author of Ref.[\onlinecite{Mila1}] found that the correlation between the two remaining spins on the Zinc-depleted triangles takes a large value of $\langle \mathbf{S}_{i}\cdot\mathbf{S}_{j}\rangle\approx -0.69$, which is very close to the theoretical limit of $-\frac{3}{4}$ for a pure spin singlet pair. According to our variational study, the corresponding spin correlation takes a value of $\langle \mathbf{S}_{i}\cdot\mathbf{S}_{j}\rangle\approx -0.67$ in the optimized $U(1)$ state and $\langle \mathbf{S}_{i}\cdot\mathbf{S}_{j}\rangle\approx -0.62$ in the optimized $Z_{2}$ state. Such a peculiar behavior is a direct consequence of the strongly frustrated nature of the antiferromagnetic coupling on the Kagome lattice and is responsible for the observed strong suppression of the local spin susceptibility around the doped Zinc ions in NMR experiment\cite{Mendels1}. We will return to this point in the next subsection.

\begin{figure}
\includegraphics[width=9cm]{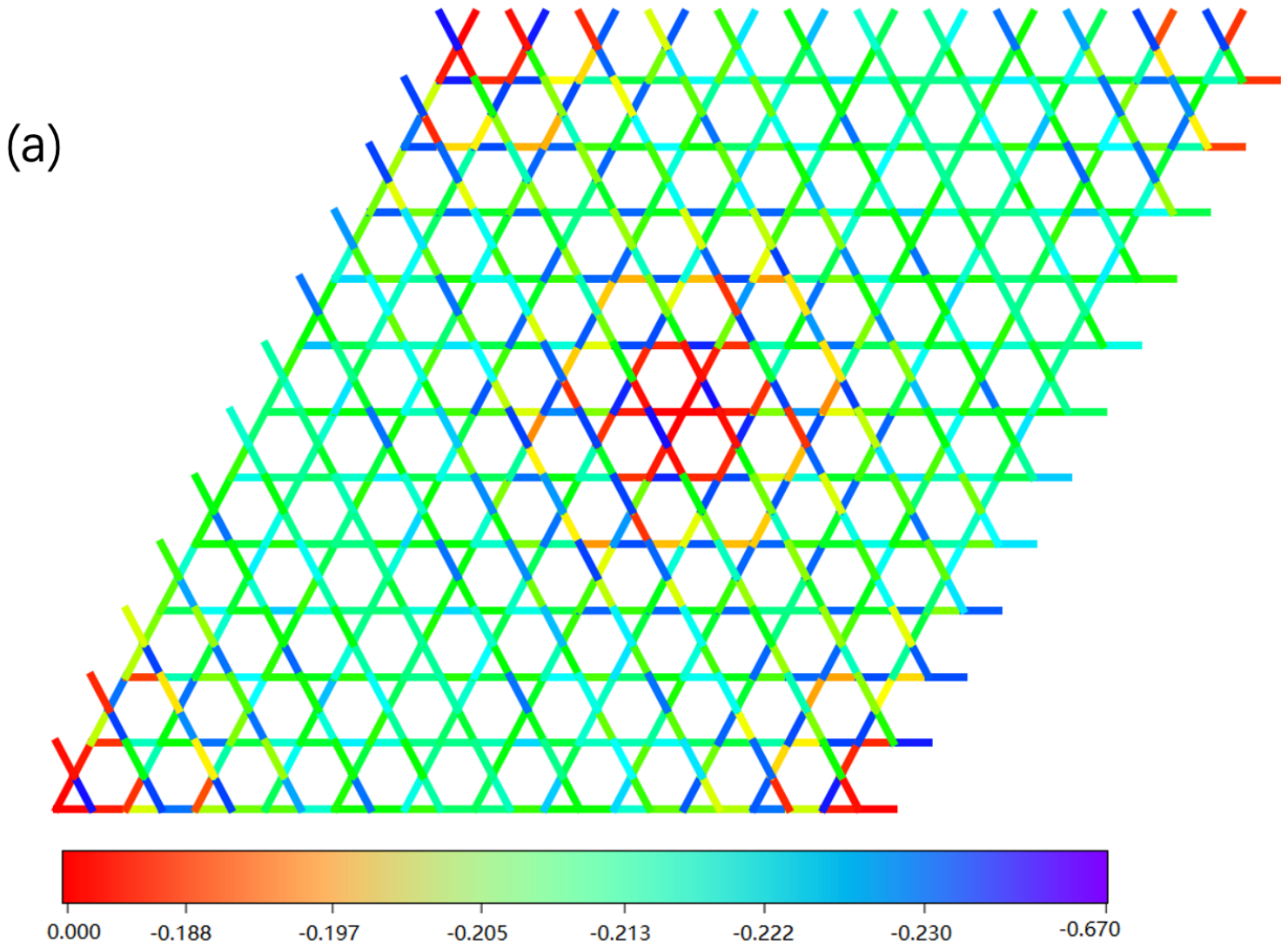}
\includegraphics[width=8cm]{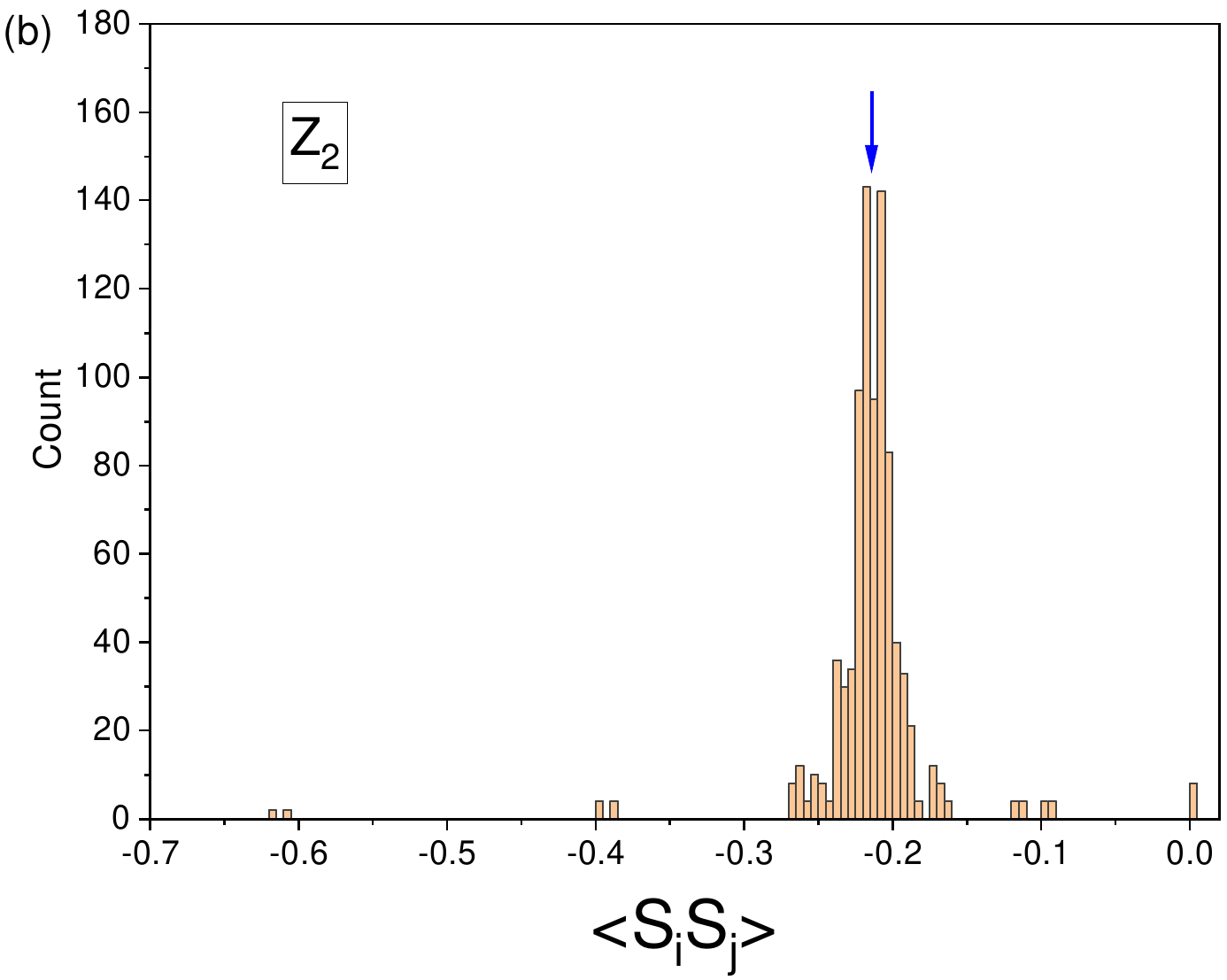}
\caption{(a)The local spin correlation pattern in the optimized $Z_{2}$ state. Note that the color scale is not uniform. (b)The distribution of the spin correlation between nearest neighboring sites. The blue arrow marks the value of nearest neighboring spin correlation in the pure spin-$\frac{1}{2}$ KAFH model.}
\end{figure}     

From the local spin correlation we can calculate the center of gravity of the local spin fluctuation spectrum, namely $\langle \omega\rangle_{i,i}$. Fig.6 plots the spatial distribution of $\langle \omega\rangle_{i,i}$ in both the optimized $U(1)$ and $Z_{2}$ state.  The spatial inhomogeneity of $\langle \omega\rangle_{i,i}$ in the $Z_{2}$ state is smaller than that in the $U(1)$ state and are both of the order of $10\%$ of their background values. This is too small to account for the observed inhomogeneity in the spin relaxation rate.

\begin{figure}
\includegraphics[width=8cm]{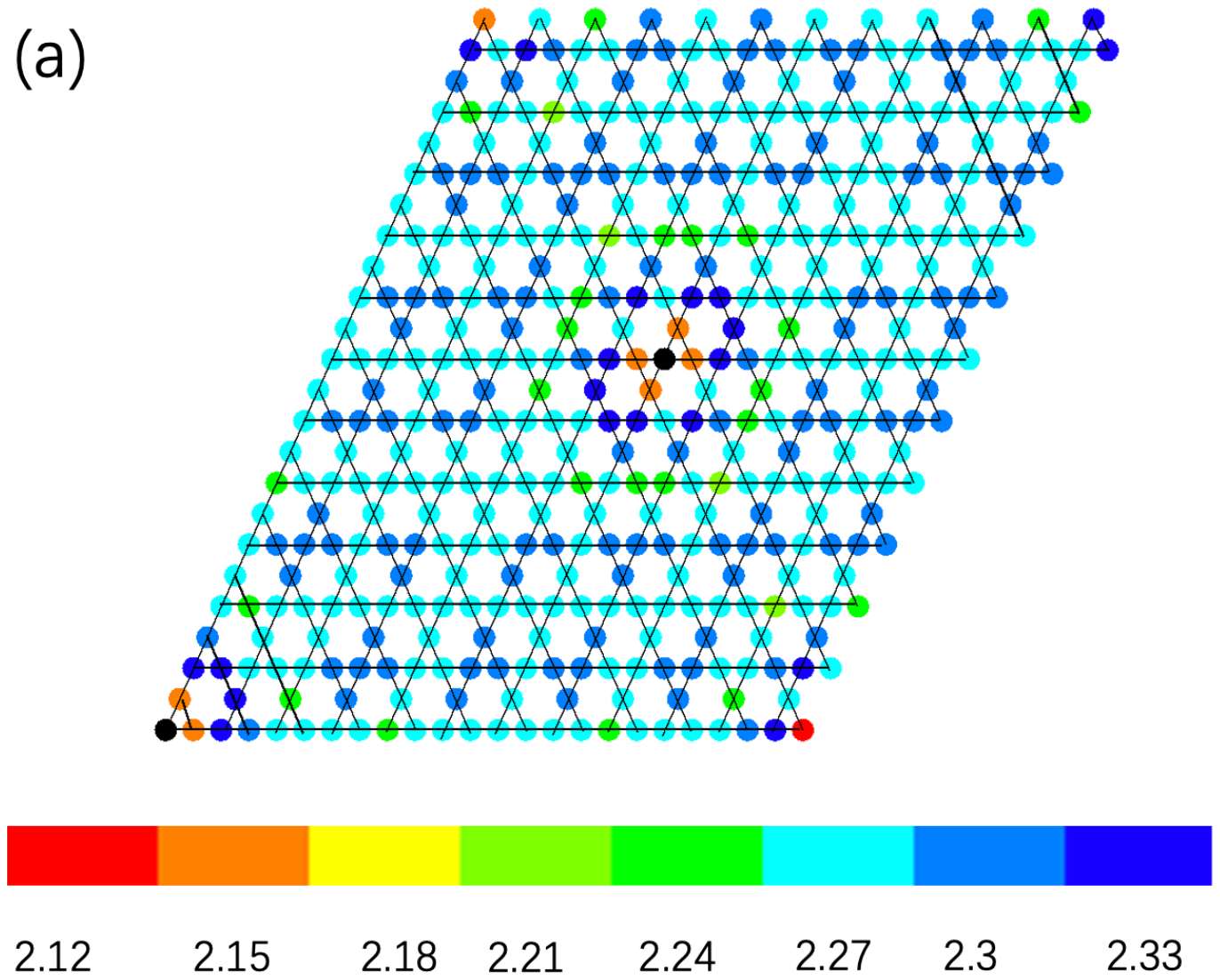}
\includegraphics[width=8cm]{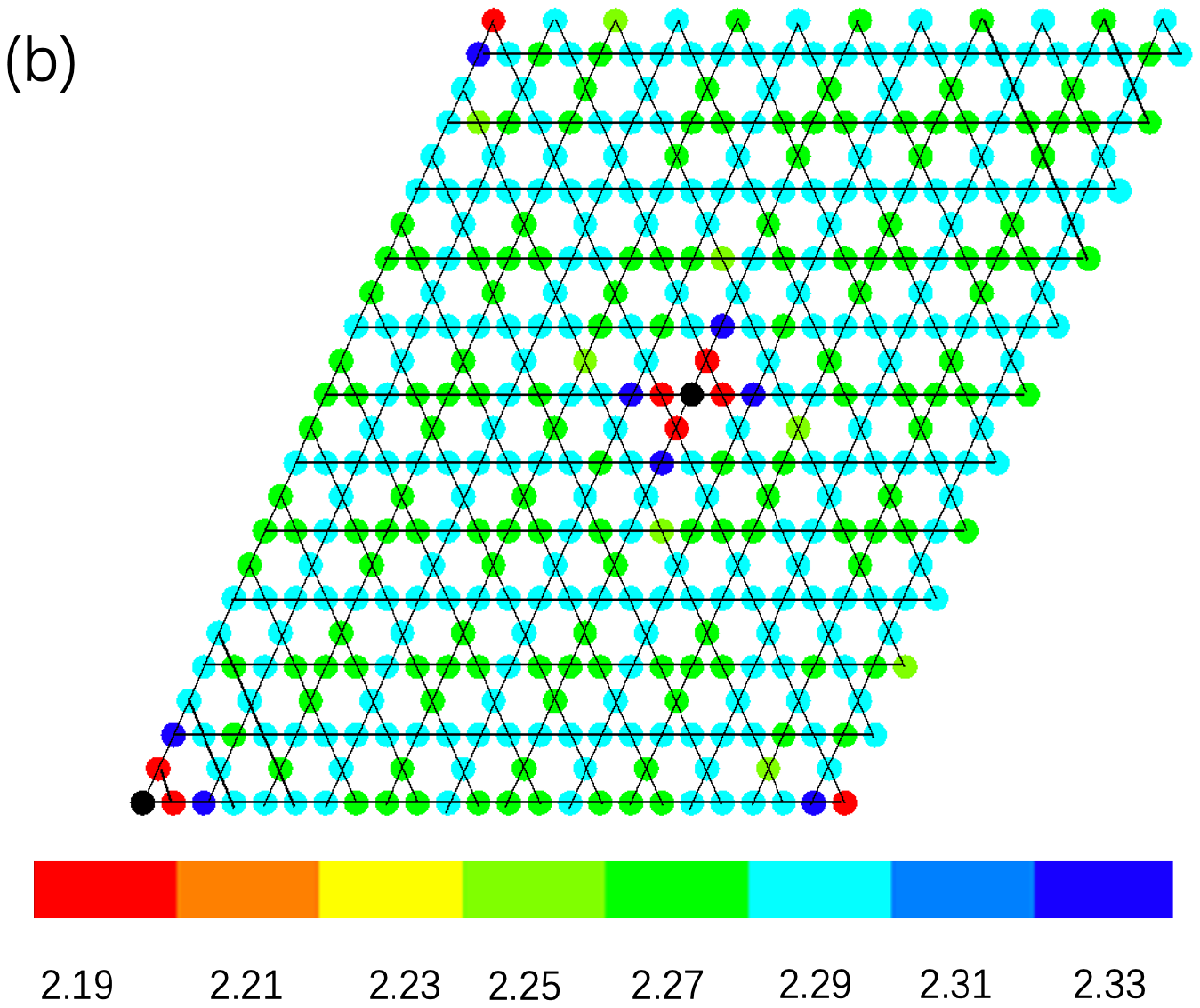}
\caption{The spatial distribution of $\langle \omega\rangle_{i,i}$, the center of gravity of the local spin fluctuation spectrum in the (a) $U(1)$ and the (b) $Z_{2}$ state. The black dots mark the position of the zinc ions.}
\end{figure}      

\subsection{Spatial inhomogeneity in the local spin susceptibility}
\begin{figure}
\includegraphics[width=8cm]{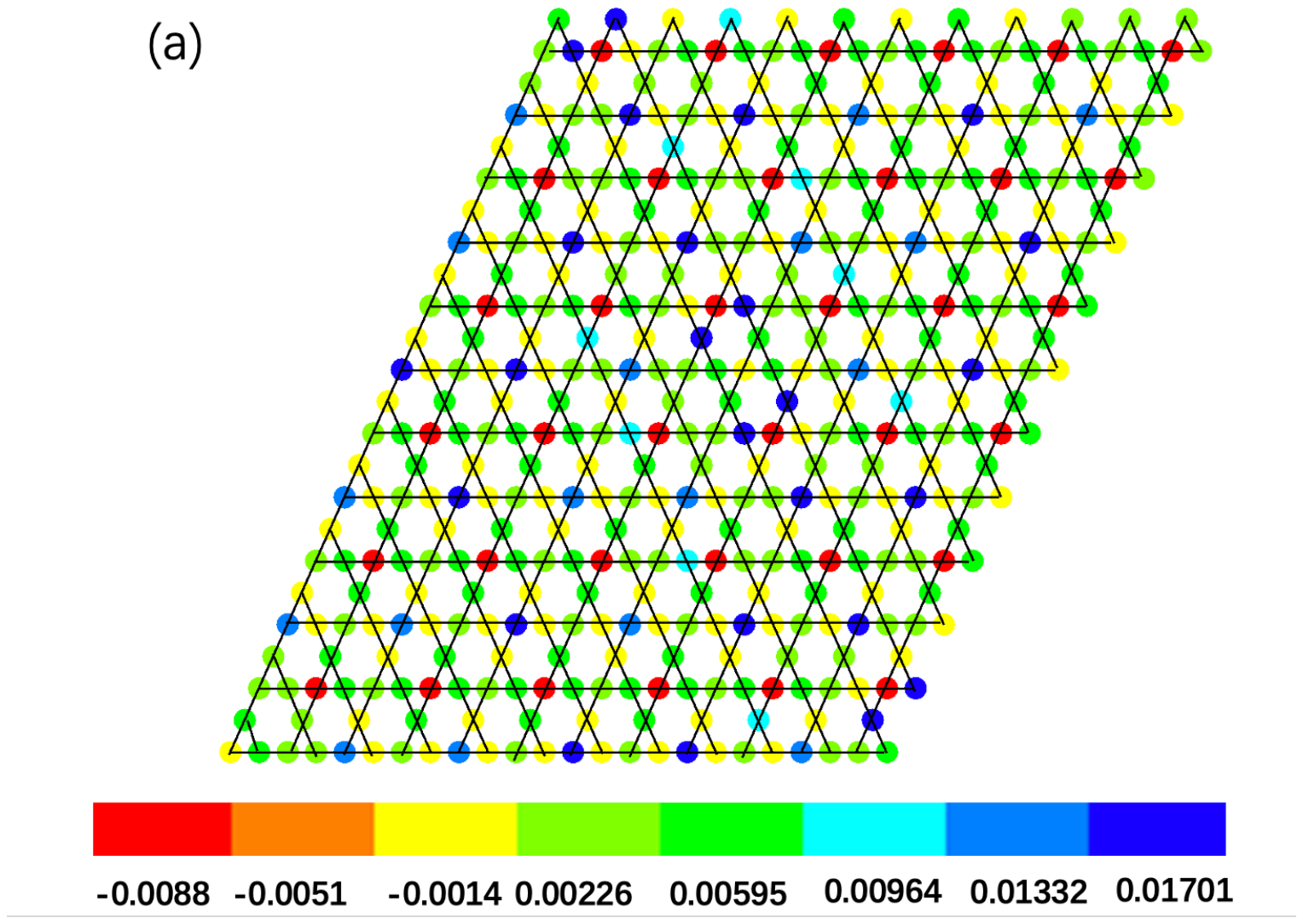}
\includegraphics[width=8cm]{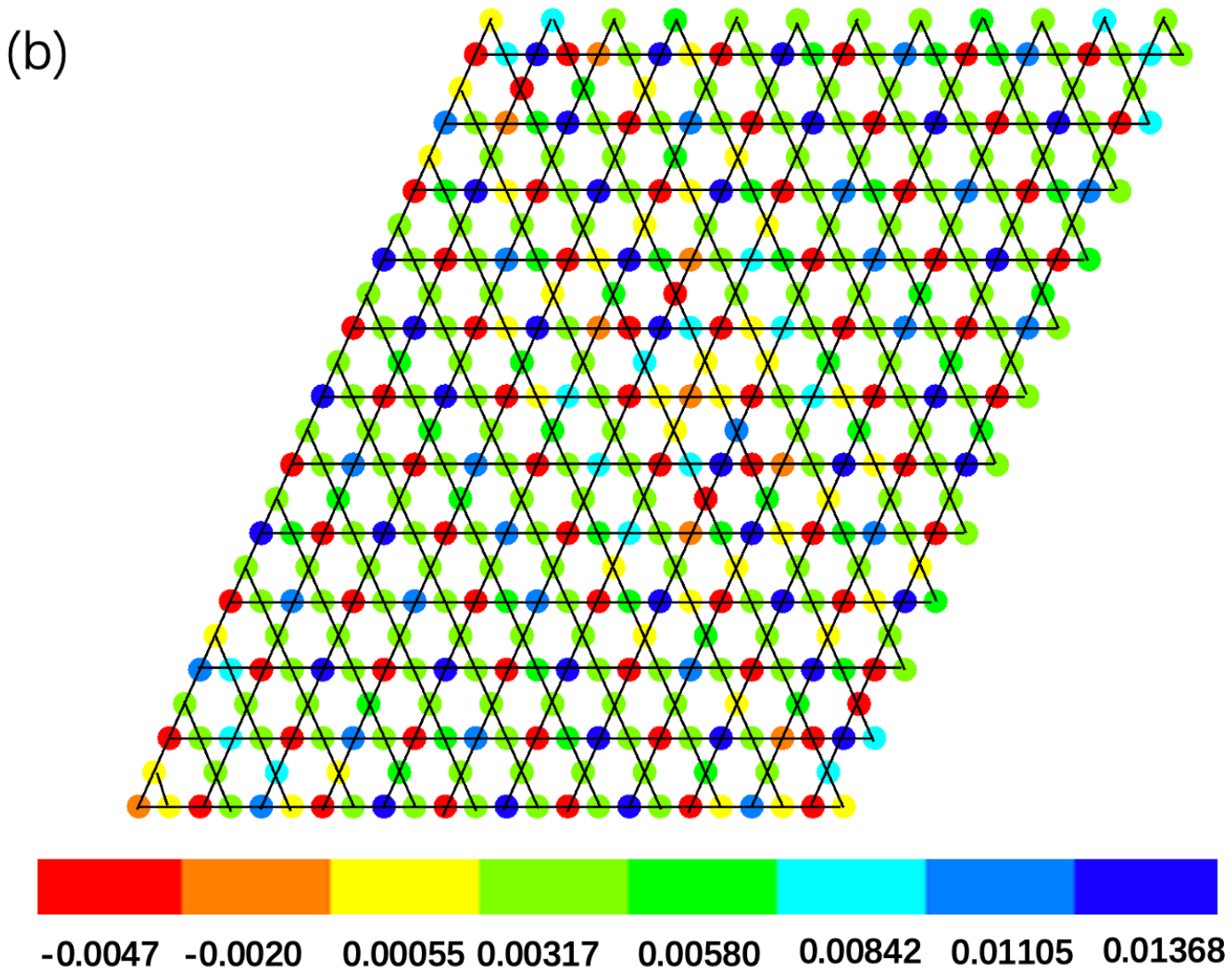}
\includegraphics[width=7cm]{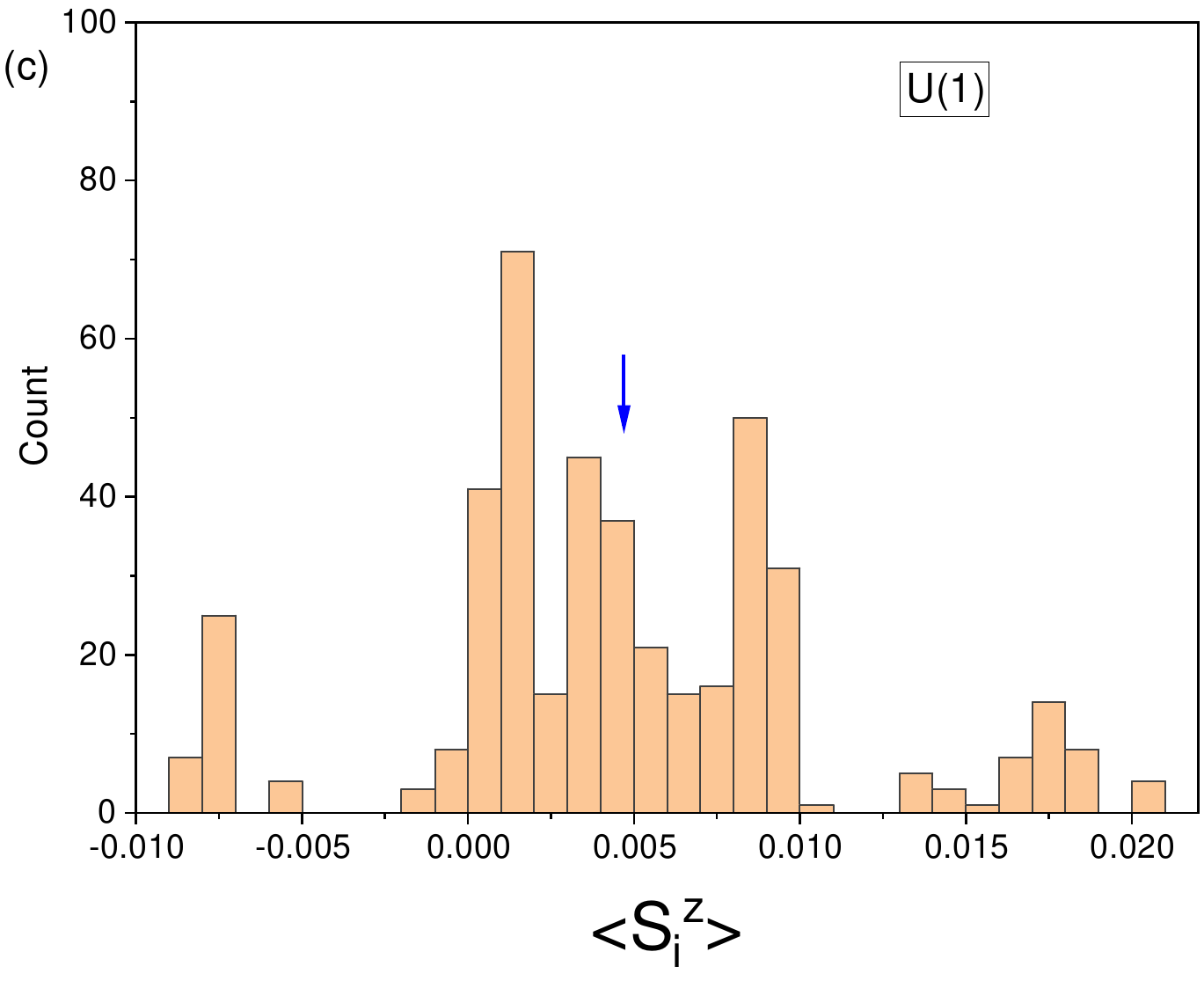}
\includegraphics[width=7cm]{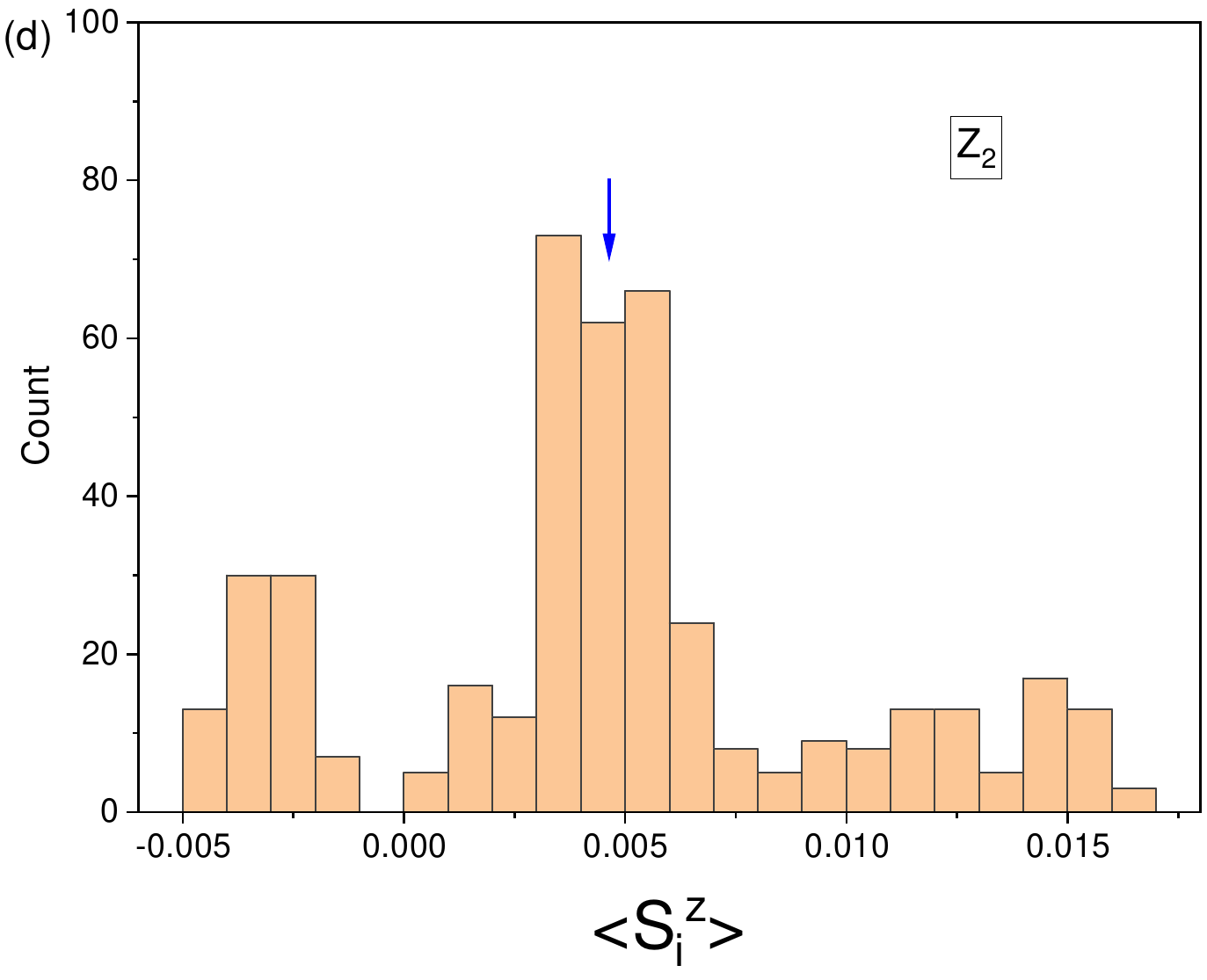}
\caption{The spatial distribution of magnetization in the optimized $U(1)$(a) and $Z_{2}$(b) state. (c) and (d) show the histogram of magnetization on the lattice in the $U(1)$ and $Z_{2}$ state. The blue arrow marks the value of the magnetization we expect when it is spatially uniform.}
\end{figure}   

Fig.7 shows the distribution of the magnetization calculated from the optimized $U(1)$ and $Z_{2}$ state. Both distributions exhibit strong oscillation behavior, but the pattern of the oscillation are different. In fact, we find that the oscillation pattern of the magnetization is not universal. For example, if we insert a gauge flux of $\pi$ in the hole surrounded by the circumference along the $\mathbf{a}_{2}$ direction in the $Z_{2}$ RVB mean field ansatz(or, change the boundary condition of the mean field ansatz along the $\mathbf{a}_{2}$ direction from periodic to anti-periodic), the oscillation pattern of the magnetization will be totally changed(as illustrated in Fig.8), although the increase in the variational energy is tiny($\Delta E\approx 0.1$).

\begin{figure}
\includegraphics[width=8cm]{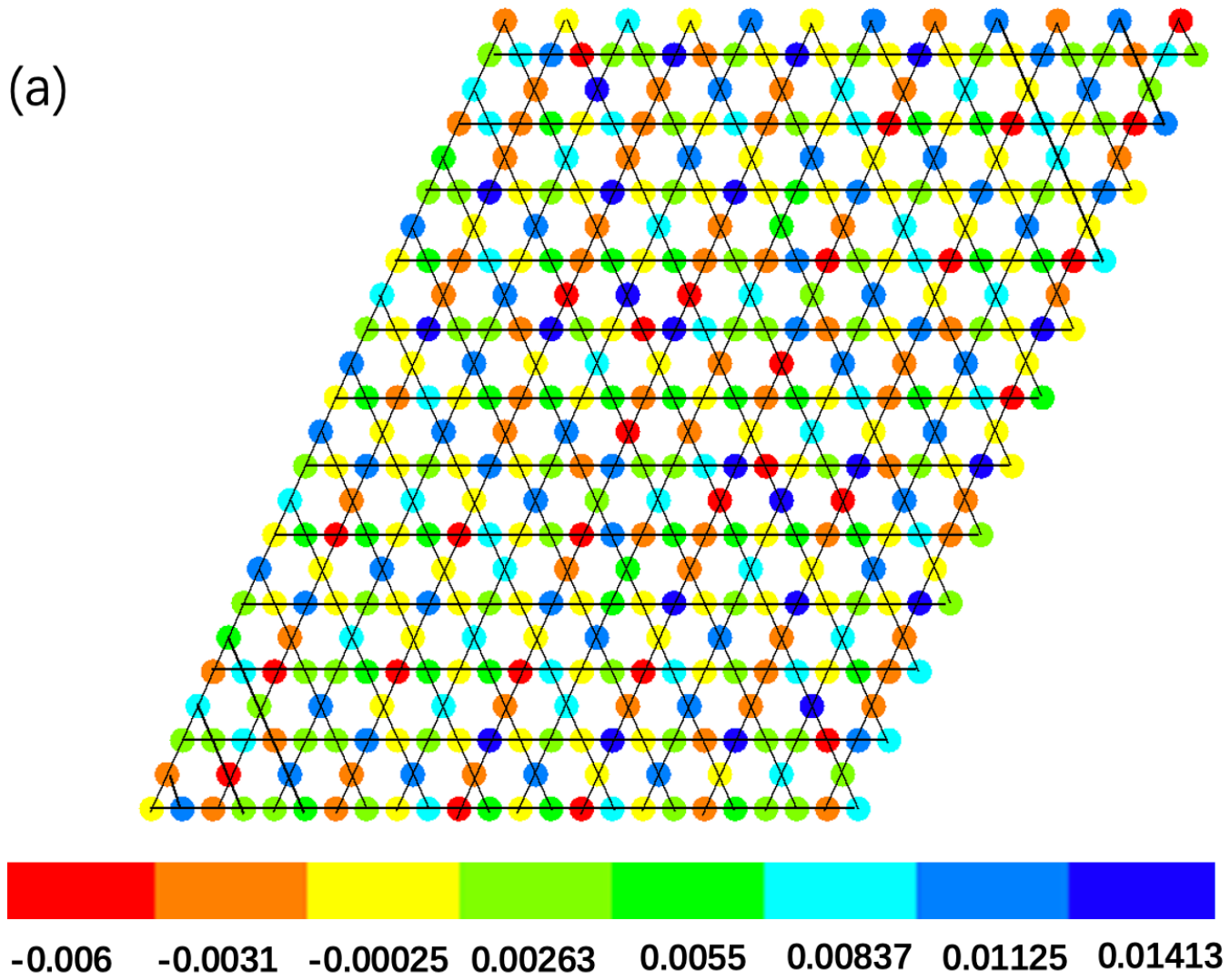}
\includegraphics[width=7cm]{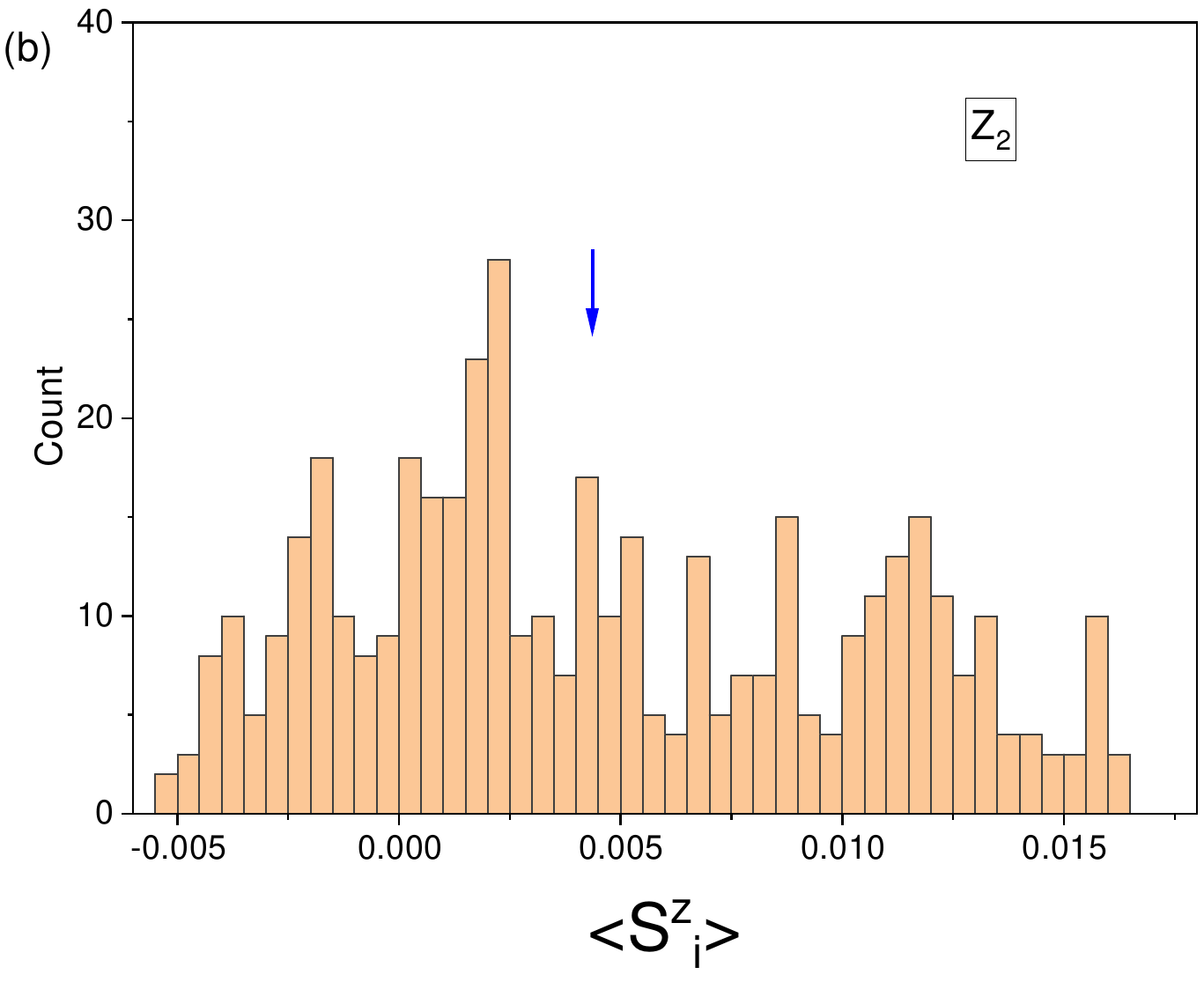}
\caption{The spatial distribution of magnetization in the $Z_{2}$ state when a gauge flux of $\pi$ is inserted in the hole surrounded by the circumference along the $\mathbf{a}_{2}$ direction. (b) shows the histogram of magnetization on the lattice in this state. The blue arrow marks the value of the magnetization we expect when it is spatially uniform.}
\end{figure}   

To see more clearly the structure of the oscillation pattern in the magnetization, we have performed Fourier transformation on the magnetization $\langle S^{z}_{i}\rangle$ for each of the three sublattices of the Kagome lattice, which are defined as
\begin{equation}
S_{\mu}(\mathbf{q})=\sum_{i\in\mu}e^{i\mathbf{q}\cdot \mathbf{R}_{i} } \langle S^{z}_{i}\rangle
\end{equation}
in which $\mu=1,2,3$ is the sublattice index, $\mathbf{R}_{i}$ is the cell index of site $i$. $\mathbf{q}=q_{1}\mathbf{b}_{1}+q_{2}\mathbf{q}_{2}$ is a wave vector in the first Brillouin zone of the Kagome lattice, with $\mathbf{b}_{1}$ and $\mathbf{b}_{2}$ the reciprocal vectors.  
The strength of the Fourier component is shown in Fig.9 for each of the three sublattices. The oscillation in the $U(1)$ state is seen to have a dominant wave vector of $\frac{1}{2}\mathbf{b}_{1}$ and $\frac{1}{2}\mathbf{b}_{2}$. The oscillation in the $Z_{2}$ state has a dominant wave vector of $\frac{1}{2}(\mathbf{b}_{1}+\mathbf{b}_{2})$. When we trap a gauge flux of $\pi$ in the hole surrounded by the circumference along the $\mathbf{a}_{2}$ direction in the $Z_{2}$ RVB ansatz, the oscillation of magnetization at both $\frac{1}{2}\mathbf{b}_{1,2}$ and $\frac{1}{2}(\mathbf{b}_{1}+\mathbf{b}_{2})$ appears. Thus, while the detailed oscillation pattern is not universal, the dominating wave vector of the magnetization oscillation can only be $\frac{1}{2}\mathbf{b}_{1}$ or its symmetry equivalents. We note that these wave vectors just correspond to the momentum difference between the two Dirac nodes of the $U(1)$ Dirac spin liquid state of the pure spin-$\frac{1}{2}$ KAFH model(see the illustration presented in Fig.10). Such a correspondence is in some sense surprising, since the translational symmetry is severely broken in the optimized RVB state. We note however that the gauge flux pattern of the $U(1)$ Dirac spin liquid state is retained in the optimized RVB state. 

\begin{figure}
\includegraphics[width=10cm]{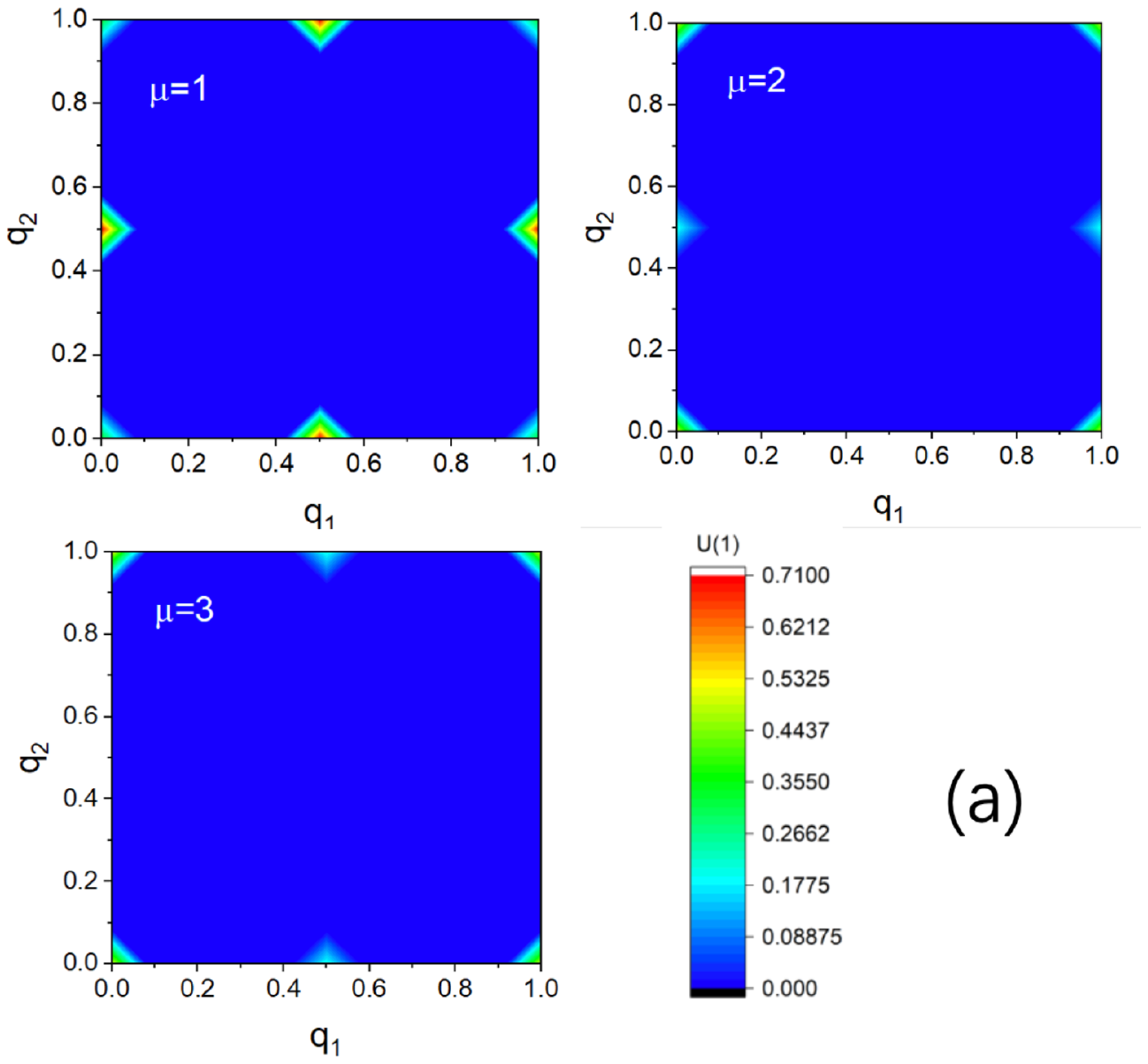}
\includegraphics[width=10cm]{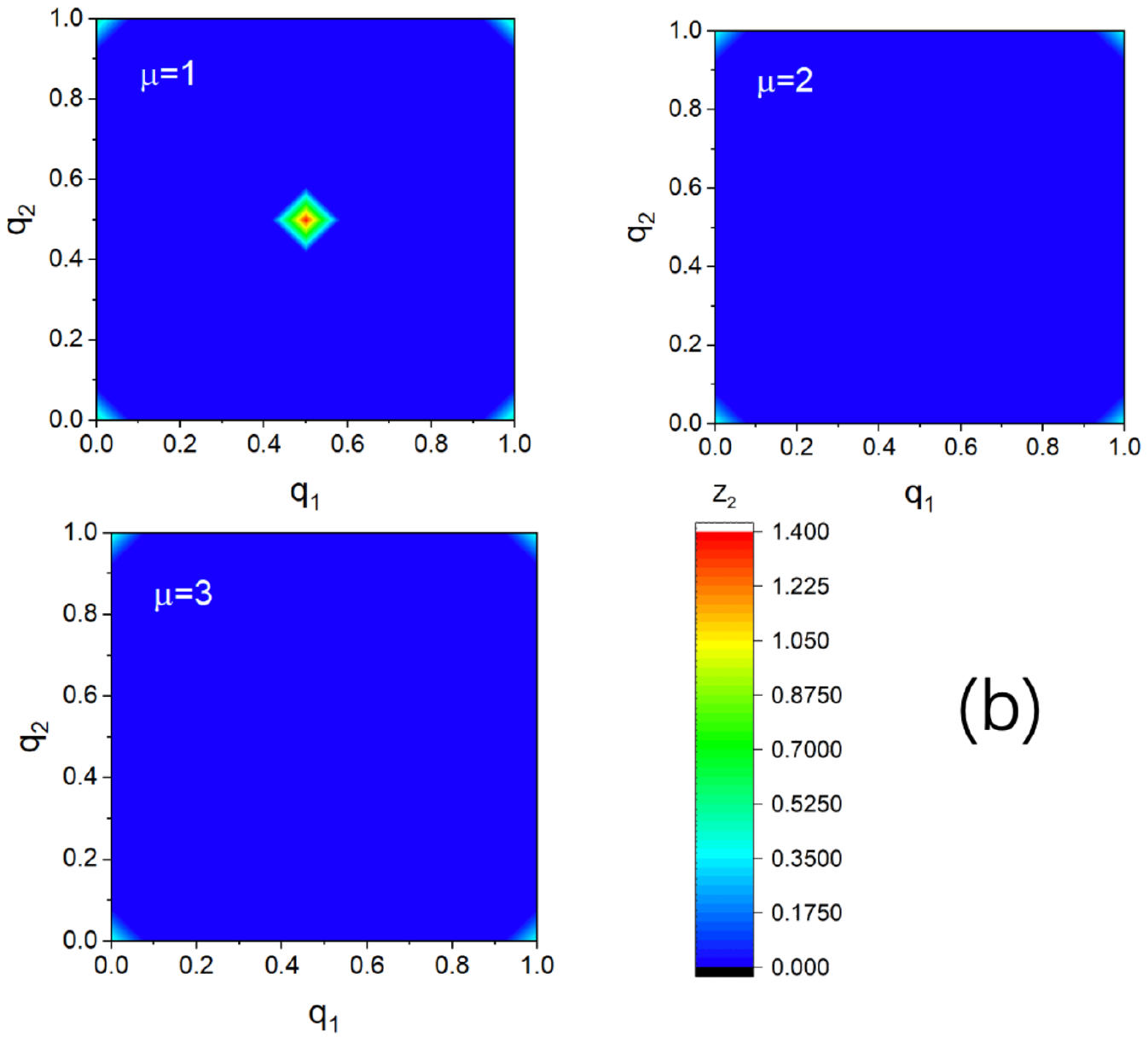}
\includegraphics[width=10cm]{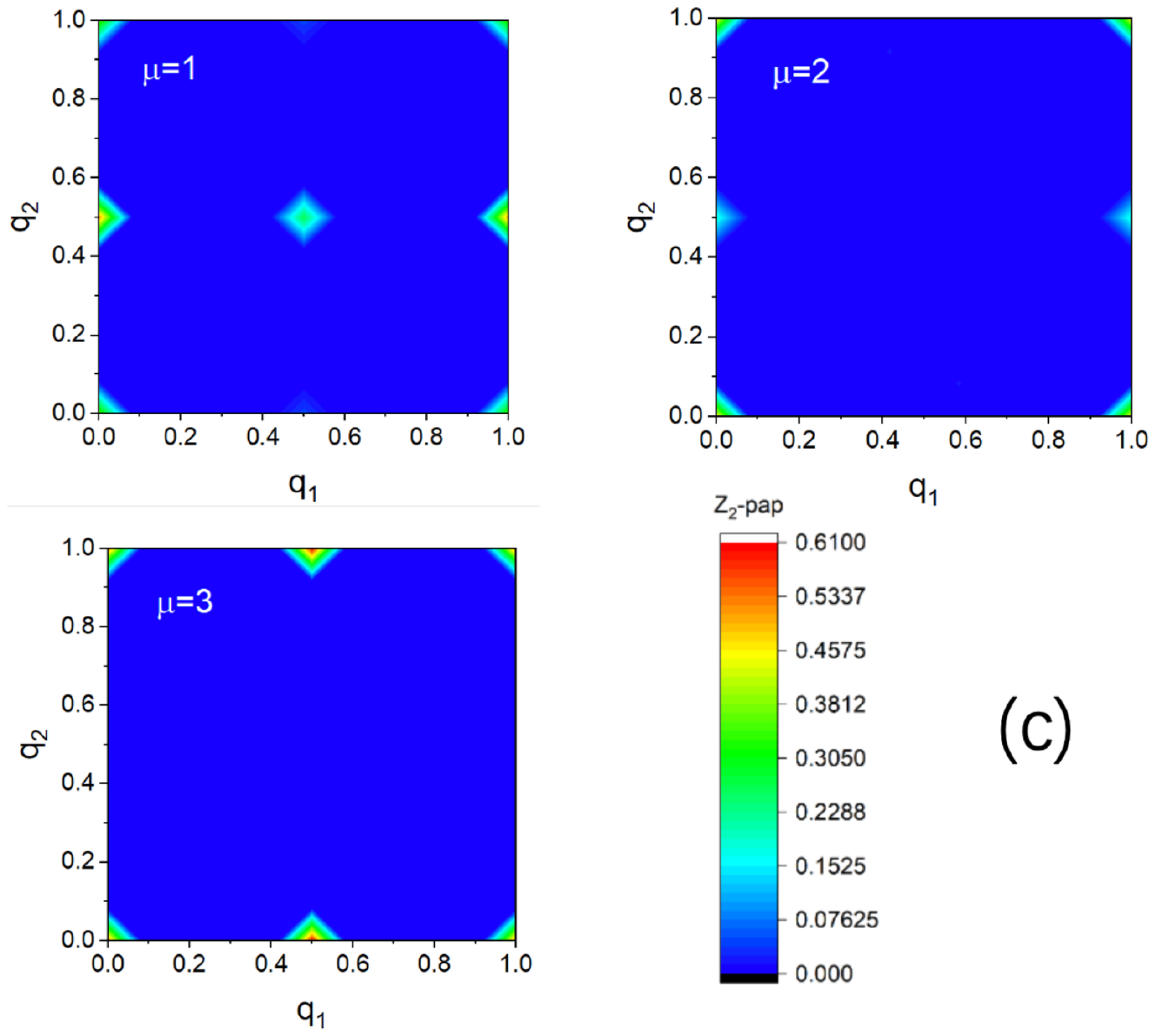}
\caption{The Fourier component of the magnetization in the three sublattices of the Kagome lattice, namely $|S_{\mu}(\mathbf{q})|^{2}$, for (a) the $U(1)$ state, (b) the $Z_{2}$ state and (c) the $Z_{2}$ state trapped with a gauge flux of $\pi$ along the direction of $\mathbf{a}_{2}$. Here $\mu=1,2,3$ is the sublattice index. Note that finer details in $|S_{\mu}(\mathbf{q})|^{2}$ are overwhelmed by the dominating peaks in these figures.}
\end{figure}   

\begin{figure}
\includegraphics[width=9cm]{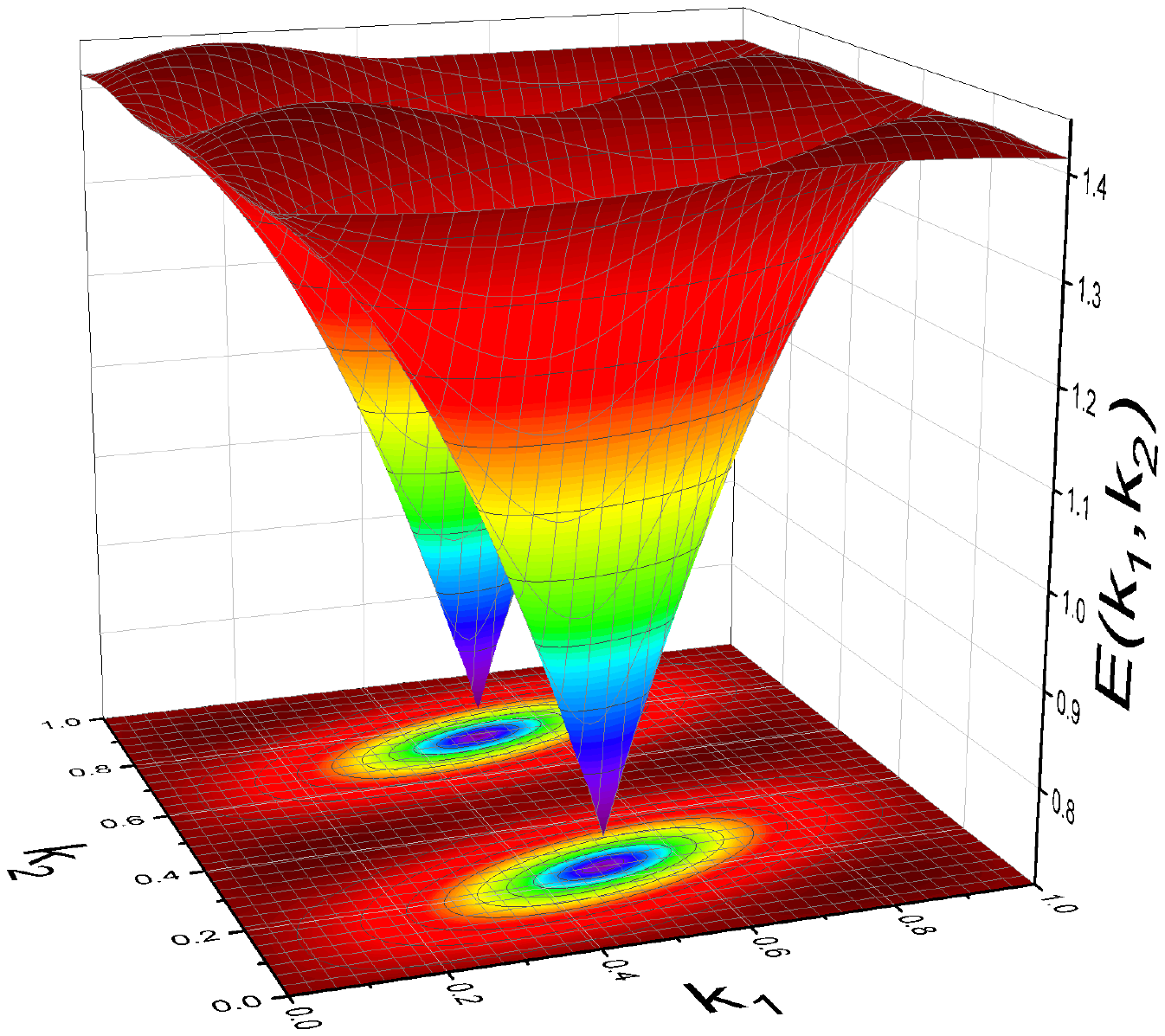}
\caption{Illustration of the Dirac nodes in the mean field dispersion of the $U(1)$ Dirac spin liquid state for the pure spin-$\frac{1}{2} $ KAFH model. The wave vector component $k_{1,2}$ is measured in unit of the reciprocal vectors of the reduced Brillouin zone for the spinon in the  $U(1)$ Dirac spin liquid state, namely $\mathbf{k}=k_{1}\frac{\mathbf{b}_{1}}{2}+k_{2}\mathbf{b}_{2}$.}
\end{figure}   

Different from the detailed pattern of the oscillation, we find that the strength of the oscillation in the magnetization is relatively insensitive to the choice of form of the mean field ansatz or its boundary condition. To measure quantitively the strength of the oscillation, we define the mean square root of fluctuation in the magnetization as
\begin{equation}
\Delta \langle S_{i}^{z}\rangle=\sqrt{\overline{\langle S_{i}^{z}\rangle^{2}}-(\overline{\langle S_{i}^{z}\rangle})^{2}}
\end{equation} 
in which
\begin{eqnarray}
\overline{\langle S_{i}^{z}\rangle^{2}}&=&\frac{1}{N}\sum_{i}\langle S_{i}^{z}\rangle^{2}\nonumber\\
\overline{\langle S_{i}^{z}\rangle}&=&\frac{1}{N} \sum_{i} \langle S_{i}^{z}\rangle
\end{eqnarray}
It is found that in the optimized $U(1)$ state $\Delta \langle S_{i}^{z}\rangle\approx 1.3\times \overline{\langle S_{i}^{z}\rangle}$. In the $Z_{2}$ state, $\Delta \langle S_{i}^{z}\rangle$ is slightly smaller and $\Delta \langle S_{i}^{z}\rangle\approx 1.1\times  \overline{\langle S_{i}^{z}\rangle}$. When we trap a gauge flux of $\pi$ along the $\mathbf{a}_{2}$ direction in the $Z_{2}$ RVB state, $\Delta \langle S_{i}^{z}\rangle$ increases slightly to $\Delta \langle S_{i}^{z}\rangle\approx 1.2 \times \overline{\langle S_{i}^{z}\rangle}$. 

It should be noted that strong spatial modulation of the magnetization has already been noticed in earlier studies\cite{Mila1,Motrunich}. In particular, it is found by exact diagonalization on small clusters\cite{Mila1} that the magnetization is strongly enhanced on sites far from the Zinc site as a result of the interference between different paths on the periodic cluster they used. The interference effect is so strong that even the direction of the magnetization can be reversed. The reversion of magnetization is also reported in a variational study\cite{Motrunich} in which the $U(1)$ Dirac spin liquid ansatz is simply truncated on the bond connected to the impurity site. Exactly the same trend is observed in our variational study with systematic optimization of the RVB ansatz. In addition, it is found that the local spin susceptibility on the Zinc-depleted triangles is strongly suppressed\cite{Mila1}, consistent with the observation of strong spin singlet pairing between the two remaining spins on such triangles. Such a unique behavior of the Kagome antiferromagnet\cite{Mendels1} is well reproduced in our variational calculation.    

Finally, through refined optimization with the BFGS algorithm\cite{Tao5}, we find that the relaxation of the RVB order parameter in the magnetized state $|Mag\rangle$ is negligibly small. At the same time, we find that the excitation of a pair of spinons in $|Mag\rangle$ only results in an increase in energy of about $0.07$ in both the $U(1)$ and the $Z_{2}$ state. This implies that the pair of excited spinons are almost unbounded with each other in the variational ground state. A natural interpretation of such a peculiar behavior is that each of the pair of excited spinons is actually a free moment released by the doped Zinc ions.

From these results, we conclude that a large spatial inhomogeneity in the magnetization is a robust prediction of the variational theory based on the Fermionic RVB picture. The oscillation in the magnetization has a dominating wave vector of $\frac{1}{2}\mathbf{b}_{1}$ or its symmetry equivalents, the momentum difference between the two Dirac nodes of the $U(1)$ Dirac spin liquid state of the pure spin-$\frac{1}{2}$ KAFH model. Such a peculiar behavior in the magnetization is very likely related to the release of free moment by the doped Zinc ions. These free moments will not only contribute to the spatial inhomogeneity of the Knight shift, but will certainly contribute to that of the spin relaxation rate.

\section{Conclusions and Discussions}

From either semiclassical argument or the RVB picture, one expect local impurity to play an important role in a highly frustrated quantum magnet. Recent NMR studies on the Kagome materials do suggest strong relevance of disorder effect in these highly frustrated quantum magnet systems.  With these considerations in mind, we have performed a systematic variational Monte Carlo study on the ground state and magnetic response of a Zinc-doped spin-$\frac{1}{2}$ KAFH model within the RVB picture, in which the doped Zinc ions act as vacancies in the Kagome plane. 

We find that the Zinc ions will significantly reorganize the spin correlation pattern around them, resulting in strong spatial modulation in the local spin correlation. The center of gravity of the local spin fluctuation spectrum, which acts as a rough description of the local dynamical behavior of the system, also exhibits moderate spatial inhomogeneity of about $10\%$ of the background value. 

The most significant effect of the doped Zinc ion manifests itself in the strong spatial inhomogeneity in the magnetic susceptibility of the system at low field. We find that the magnetization of the Zinc-doped system exhibits strong spatial oscillation with a dominant wave vector of $\frac{1}{2}\mathbf{b}_{1}$ or its symmetry equivalents. The amplitude of such oscillation is found to be even larger than the average magnetization. We find that this is a robust prediction of the RVB picture and is insensitive to the choice of the form the RVB mean field ansatz or its boundary condition and the relaxation of the RVB structure in the magnetized state.

In our construction of the magnetized state $|Mag\rangle$, we have excited a pair of spinons above the RVB mean field ground state. These pair of spinons are found to have very small excitation energy in the optimized RVB state. We suggest to interpret each of the pair of excited spinons as a free moment released by the doped Zinc ions. While a free moment is naturally expected when we introduce a lattice vacancy in a RVB background, since the spin involved in the valence bond with the spin previously occupying the vacancy site is now released, what is extremely unusual here is that the magnetic response of such a free moment is so extended in space. These free moments will not only contribute to the spatial inhomogeneity of the Knight shift, but will also contribute to that of the spin relaxation rate. 

We find that the dominant wave vectors in the oscillation of the magnetization are just the momentum difference between the two Dirac nodes of the $U(1)$ Dirac spin liquid state on the Kagome lattice. This may not be just a coincidence and may imply the relevance of the nodal spinon in the magnetization process of the Zinc-doped system, even though the translational symmetry is severely broken. We propose that the oscillation in the magnetization in the Zinc-doped system may provide a smoking gun evidence for the Dirac node of the $U(1)$ Dirac spin liquid state on the Kagome lattice.   

The strong relevance of disorder effect in highly frustrated quantum magnet is generally expected and surely not restricted to the Kagome magnet. We thus expect that the study presented in this paper may stimulate further study of disorder effect in other quantum spin liquid candidate systems, from both theoretical and experimental perspectives. In particular, it is interesting to verify the predictions of our variational study on the Zinc-doped spin-$\frac{1}{2}$ KAFH model using other numerical approaches or NMR measurement on Kagome materials with deliberately doped Zinc ions in the Kagome plane.

A related issue of conceptual importance is the fate of the topological order, or more generally, the quantum order and the corresponding fractionalized excitation in the presence of disorder effect for such highly frustrated quantum magnets. While it is generally believed that the topological order is robust against small local perturbation, it is totally unclear if the Zinc doping effect in a RVB background can be taken as a local  perturbation, since the RVB structure itself is an emergent structure when frustrated couplings are delicately balanced, rather than a robust built-in feature in the model Hamiltonian. The power of the doped Zinc ions to release free spinons, which is certainly an nonlocal excitations, also implies that it can not be taken as a local perturbation. It is thus interesting to ask if there is any qualitative distinction in the Zinc-doped system between the more conventional valence bond glass phase(or the random singlet phase) and the genuine spin liquid state with topological order or quantum order.

\begin{acknowledgments}
We acknowledge the support from the National Natural Science Foundation of China(Grant No.12274457). We also thank Pranay Patil, Philippe Mendels and Fr\'ed\'eric Mila for their valuable comments.
\end{acknowledgments}

\end{document}